\documentclass[useAMS,usenatbib]{mn2e}
\usepackage{graphicx}
\usepackage{latexsym}
\usepackage{amssymb}

\title{The similarity of the stellar mass fractions of galaxy groups and
clusters}

\author[J. M. Budzynski et
al.]{J.~M.~Budzynski$^1$\thanks{E-mail:koposov@ast.cam.ac.uk,i.g.mccarthy@ljmu.ac.uk}, S.~E.~Koposov$^{1,2}$, I.~G.~McCarthy$^{3}$, V.~Belokurov$^1$ \\
$^1$Institute of Astronomy, Madingley Road, Cambridge CB3 0HA \\
$^2$Moscow MV Lomonosov State University,
Sternberg Astronomical Institute, Moscow
119992, Russia\\
$^3$Astrophysics Research Institute, Liverpool John Moores University, 146 Brownlow Hill, Liverpool L3 5RF
}

\begin{document}

\pagerange{\pageref{firstpage}--\pageref{lastpage}} \pubyear{2013}

\newcommand{\dd}{\textrm{d}}
\newcommand{\rin}{R_\textrm{\tiny{in}}}
\newcommand{\OO}{\mathcal{O}}
\newcommand{\budpaper}{Budzynski et al. (2012)}

\maketitle

\label{firstpage}

\def\mnras{MNRAS}
\def\apj{ApJ}
\def\aap{A\&A}
\def\aaps{A\&A}
\def\apjl{ApJL}
\def\aj{AJ}
\def\apjs{ApJS}
\def\araa{ARA\&A}
\def\nat{Nature}
\def\pasp{PASP} 

\newcommand{\samplesize}{20\,171}
\newcommand{\xraysamplesize}{129}
\newcommand{\deprojection}{0.67}


\begin{abstract} We employ a large sample of \samplesize \ optically-selected groups and clusters at $0.15 \le z \le 0.4$ in the SDSS to investigate how the stacked stellar mass fraction varies across a wide range of total mass, $M_{500}$.  Our study improves upon previous observational studies in a number of important ways, including having a much larger sample size, an explicit inclusion of the intracluster light (ICL) component, and a thorough examination of the accuracy of our total mass estimates via comparisons to simulations and weak lensing observations.  
We find that the stellar mass fraction depends only weakly on total mass and that the contribution of ICL to the total stellar mass fraction is significant (typically 20-40 per cent).  Both of these findings are in excellent accordance with the predictions of cosmological simulations.  Under the assumption of a Chabrier (Salpeter) IMF, the derived star formation efficiency ($f_{star}/f_b$, where $f_b\equiv\Omega_b/\Omega_m$) is relatively low at 8 per cent (14 per cent) and is consistent with the global star formation efficiency of semi-analytic models that reproduce the galaxy stellar mass function.  When our measured stellar mass fractions are combined with the observed relation between hot gas mass fraction and total mass from X-ray observations, our results imply that galaxy groups have significantly lower baryon fractions than massive clusters.  Ejection of gas due to energetic AGN feedback (most likely at high redshift) provides a plausible mechanism for explaining the trends we observe. 
\end{abstract}


\section{Introduction} 

X-ray studies of galaxy groups and clusters show that the fractional contribution of the hot gas ($f_{\rm gas}=M_{\rm gas}/M_{500}$) to total mass of the system within relatively large radii of $r_{500}$\footnote{Defined as the radius that encloses a mean density that is 500 times the critical density of the Universe.  For concentrations typical of massive galaxy clusters, $r_{500} \approx 0.65 r_{200}$.} increases with increasing total mass \citep{1999MNRAS.305..631A, 1999ApJ...517..627M, 2000A&A...361..429R, 2003ApJ...591..749L, 2006ApJ...640..691V, 2007MNRAS.377.1457M,2009ApJ...693.1142S,2009A&A...498..361P}.  The most recent results suggest $f_{\rm gas} \propto M_{500}^{\approx 0.13}$ \citep{2009ApJ...693.1142S}.  The origin of this trend in $f_{\rm gas}$ with system mass has not yet been fully elucidated.  Cosmological simulations that lack efficient feedback predict that within $r_{500}$ the total baryon fraction of groups and clusters ought to be close to the universal baryon fraction $f_b \equiv \Omega_b/\Omega_m$ \citep[e.g.][]{1999ApJ...525..554F,2005ApJ...625..588K,2006MNRAS.365.1021E,2007MNRAS.377...41C}.  (Analysis of the {\it WMAP} 9-yr data indicates $f_b \approx 0.165$; \citealt{2013ApJS..208...19H}.)
Therefore, the lower gas mass fractions in groups with respect to clusters may simply be indicating that the fraction of baryons locked up in stars ($f_{\rm star}$; this includes stars in satellite galaxies as well as that in a diffuse intracluster component) increases with decreasing total mass so as to compensate for the decrease in $f_{\rm gas}$.  The contribution of cold (non-X-ray-emitting) gas is generally thought to be negligible in comparison to the hot gas and stellar content of clusters.

Alternatively, if the stellar mass fraction does not rise significantly with decreasing mass, the implication is that galaxy groups are deficient in their baryon content with respect to clusters and to the universal mean.  Such a scenario is not without a physical basis - \citet{2010MNRAS.406..822M,2011MNRAS.412.1965M} have shown that the inclusion of energetic AGN feedback in cosmological simulations can indeed result in lower-than-universal baryon fractions for galaxy groups.  Indeed, by examining a large number of simulations with varying input physics (the OverWhelmingly Large Simulations; \citealp{2010MNRAS.402.1536S}), McCarthy et al.\ concluded that {\it only} those simulations which include such ejective AGN feedback are able to reproduce both the X-ray and optical properties of local galaxy groups and clusters.

While it appears that a general consensus has been reached on the (hot) gas mass fractions of clusters in the local Universe, at least of relatively massive systems, the same cannot be said of the stellar mass fractions\footnote{The stellar mass fractions of groups and clusters in `optical' studies also tend to be quoted within the aperture $r_{500}$.  The motivation for selecting this radius is two-fold: (i) for comparisons $f_{\rm gas}$ measurements made within this radius using X-ray observations, which are typically limited to $r_{500}$; and (ii) optical studies often use $M_{500}$ from X-ray observations as an estimate of the total system mass.}.  Much recent observational work has been done on estimating the stellar mass fractions of groups and clusters yielding some apparently contradictory results.  \citet{2003ApJ...591..749L} and Lin \& Mohr (2004) (hereafter collectively LM04) estimated the stellar and gas mass fractions for a sample of $\approx30$ and $\approx90$ X-ray-selected clusters, respectively, with {\it 2MASS} K-band data.  They found a weakly rising stellar mass fraction with decreasing total mass $f_{\rm star} \propto M_{500}^{-0.26 \pm 0.09}$.  The net result is a total baryon fraction that increases with total mass $f_{\rm bar} \propto M_{500}^{0.15 \pm 0.04}$.  Similar results were obtained by \citet{2011MNRAS.412..947B}, who examined stellar and hot gas content of 18 nearby, low-mass clusters selected from the 2dF Galaxy Redshift Survey and followed up with deep ground-based near-infrared (K-band) observations and {\it Chandra} and {\it XMM} X-ray observations.
\citet{2007ApJ...658..917D,2010ApJ...719..119D} stacked {\it ROSAT All Sky Survey} data of over 4000 groups and clusters from the {\it 2MASS} catalog and also found a rising baryon fraction with increasing mean system temperature (which is a good proxy for total mass).  \citet{2009ApJ...703..982G} measured the stellar mass fractions of 91 X-ray-selected groups in the {\it COSMOS} 2 deg$^2$ survey and found a steeper trend in stellar mass fraction ($f_{\rm star} \propto M_{500}^{-0.37}$) than reported by LM04.  When combined with gas mass fractions from the literature, the net result is still a mildly rising baryon fraction with increasing total mass.  More recently, however, \citet{2010MNRAS.407..263A} (hereafter A10) measured the stellar mass fractions of 52 groups and clusters in the SDSS, finding an even steeper relation of $f_{\rm star} \propto M_{200}^{-0.55 \pm 0.08}$ and concluded that the overall baryon fraction is therefore approximately independent of total system mass, although it is offset in normalisation from the {\it WMAP} universal baryon fraction by approximately 6 sigma.  The origin of the differences in the findings of these studies is not obvious, but may be attributable to differences in the assumed the stellar mass-to-light ratio ($M/L$) (and its possible variation with cluster mass, due to the change in stellar populations with system mass), differences in techniques for estimating $M_{500}$, as well as different sample selection criteria.

It is worth noting that none of the above mentioned studies were able to
directly take into account the contribution from diffuse intracluster light
(ICL), due to its low surface brightness.  Whether or not this is a serious
omission is a matter of some debate.  For example,
\citet{2005MNRAS.358..949Z} stacked SDSS imaging data of approximately 700
clusters and found that, on average, the ICL contributes only a small
fraction of the total light ($\approx 10$ per cent).  In contrast,
\citet{2007ApJ...666..147G,2005ApJ...618..195G} (hereafter G07) used deep
I-band data to study the ICL of 23 local groups and clusters and found that,
in some cases, it can actually contain the majority of the total light of
the system, particularly for low-mass groups \citep[see
also][]{2010MNRAS.403L..79M}.  G07 infer a very steep stellar mass fraction
trend of $f_{\rm star} \propto M_{500}^{-0.64 \pm 0.13}$ and, when combined
with $f_{\rm gas}$ measurements from the literature, conclude that the total
baryon fraction is independent of total system mass, though it is offset
from universal mean by several sigma.

More recently, \citet[][hereafter L12]{2012ApJ...746...95L} used multi-band
COSMOS data to infer the stellar mass fractions of a local sample of
X-ray-selected groups and low-mass clusters.  In a complementary approach,
they also derived the trend over a much wider range in system masses using a
statistical Halo Occupation Distribution (HOD) model which is determined
from a joint fit to galaxy-galaxy weak lensing, galaxy clustering, and the
stellar mass function all measured from COSMOS data.  As pointed out by L12,
the studies mentioned above inferred the total stellar mass by applying a
simple stellar $M/L$ ratio conversion.  L12 argue that applying a single
conversion factor to all galaxies will bias the estimate of $f_{\rm star}$,
since not all galaxies in groups and clusters are quiescent and furthermore
the fraction of `quenched' galaxies is known to change systematically with
total system mass \citep[e.g.][]{2006MNRAS.366....2W}.  L12 find only a weak
dependence of the stellar mass fraction on total system mass and a much
lower amplitude than found previously by \citet{2009ApJ...703..982G} and
G07, which L12 attribute to the use of inaccurate $M/L$ estimates in these
studies and/or because they assumed all galaxies were quiescent early-types. 
The implication of the L12 results is that the overall baryon fraction is a
relatively steep function of total mass.  Note that L12 did not include the
ICL contribution in their stellar mass estimates.  While they have argued
that this is not the primary reason for the difference in their results from
those of G07, it is nevertheless possible that the ICL could contribute a
non-negligible fraction of the baryon budget of groups and clusters.

The findings of L12 suggest that the differences in the way in which stellar masses are estimated may be significant and may explain the difference in the inferred relation between stellar mass fraction and system mass.  We point out that not only are there differences in the way that the stellar masses are inferred, but there are also significant differences in the way the {\it total} system mass is inferred.  For example, LM04 use a hydrostatic mass-X-ray temperature scaling relation to infer total mass, \citet{2009ApJ...703..982G} and L12 use a stacked X-ray luminosity-weak lensing mass relation, G07 use a hydrostatic mass-galaxy velocity dispersion relation, and A10 use the galaxy caustic method.  These relations have not been directly compared against one another and it is entirely plausible that part of the reported differences are attributable to differences in the total mass estimates\footnote{With the exception of A10 all of the above mentioned studies define the total mass as $M_{500}$ and are therefore meant to at least be measuring the same quantity.}.  Note that the situation is complicated by the fact that the stellar masses are measured within an aperture that is determined by the estimated total mass (e.g., $r_{500}$), rather than within a fixed physical aperture (e.g., 1 Mpc).

In the present study, we revisit the stellar mass fractions of groups and
clusters but using a somewhat different approach.  In particular, in
\citet[][hereafter B12]{2012MNRAS.423..104B}, we presented a new
optically-selected catalog of \samplesize \ high-mass groups and clusters
with $M_{500}>10^{13.7}$ $M_{\odot}$ over the redshift range $0.15 \leq z
\leq 0.4$ from the SDSS.  Using a subsample of these groups and clusters
with X-ray temperature measurements, B12 calibrated an optical
richness-hydrostatic mass relation by employing the hydrostatic
mass-temperature relation of \citet{2006ApJ...640..691V} and
\citet{2009ApJ...693.1142S}.  An advantage of using X-ray temperatures
rather than X-ray luminosity or velocity dispersion is the smaller scatter
that temperature has with total mass.  B12 argued that their richness-mass
relation must be quite accurate on average, given the excellent agreement in
normalisation between the number density profiles of satellite galaxies in
the observed groups and clusters and those in semi-analytic models of galaxy
formation (see Fig.\ 10 of that study and accompanying text).  Below we
confirm this accuracy by comparing the stacked line-of-sight velocity
dispersion of observed and simulated groups and clusters, as well as making
comparisons with stacked weak lensing observations of the maxBCG sample
\citep{2007ApJ...660..239K}.

We present two measures of the stellar mass fractions of our groups and
clusters, both based on stacking in bins of $M_{500}$.  In the first method,
we simply stack the SDSS galaxy catalog and use a statistical background
subtraction procedure.  This method neglects the contribution from an ICL
component but does allow for the possibility of varying galaxy populations
with system mass.  In the second method, we stack SDSS {\it images} and take
advantage of the drift scan nature of the SDSS to locally background
subtract.  This method implicitly includes the ICL and all galaxies below
the detection/inclusion threshold in the first method.  Thus, our work
improves upon previous studies in a number of ways: i) we have a much larger
sample size; ii) we explicitly explore the importance of varying galaxy
population and the contribution of the ICL; and iii) we test our total mass
estimates against models and stacked weak lensing observations.

The present paper is outlined as follows.  In Section 2 we briefly discuss our group and cluster sample and test the accuracy of our total mass estimates, $M_{500}$.  In Section 3 we present our analysis methods for estimating the stellar masses of the groups and clusters.  We present our main results in Section 4 and make comparisons with previous work.  Finally, in Section 5 we summarise and discuss our findings.

Throughout this work we assume a $\Lambda$CDM cosmological model with $\Omega _{\mathrm{M}}=0.27$, $\Omega _{\mathrm{\Lambda}}=0.73$, and $h=0.71$.

\begin{figure*}
\includegraphics[width=0.49\textwidth]{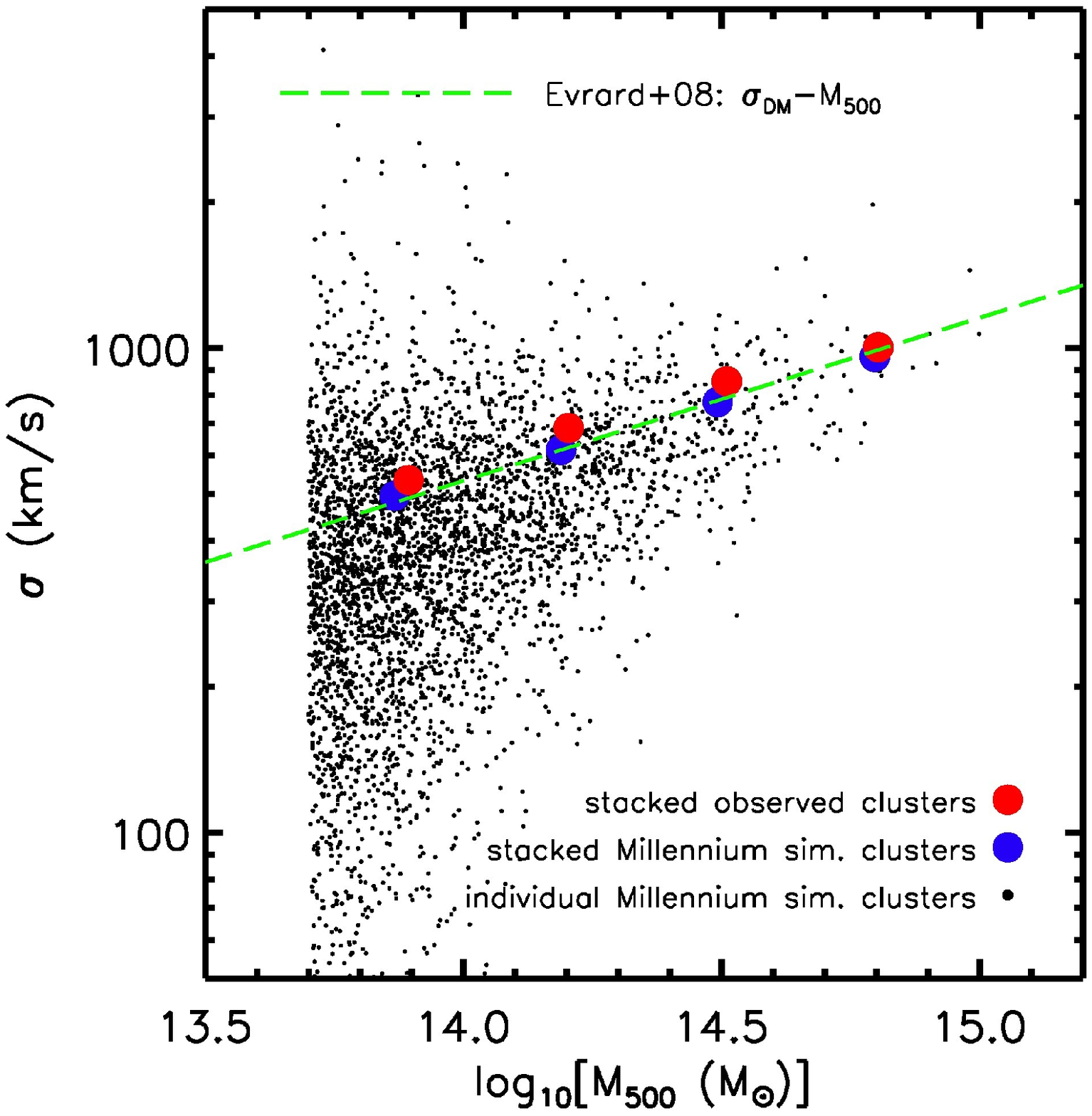}
\includegraphics[width=0.46\textwidth]{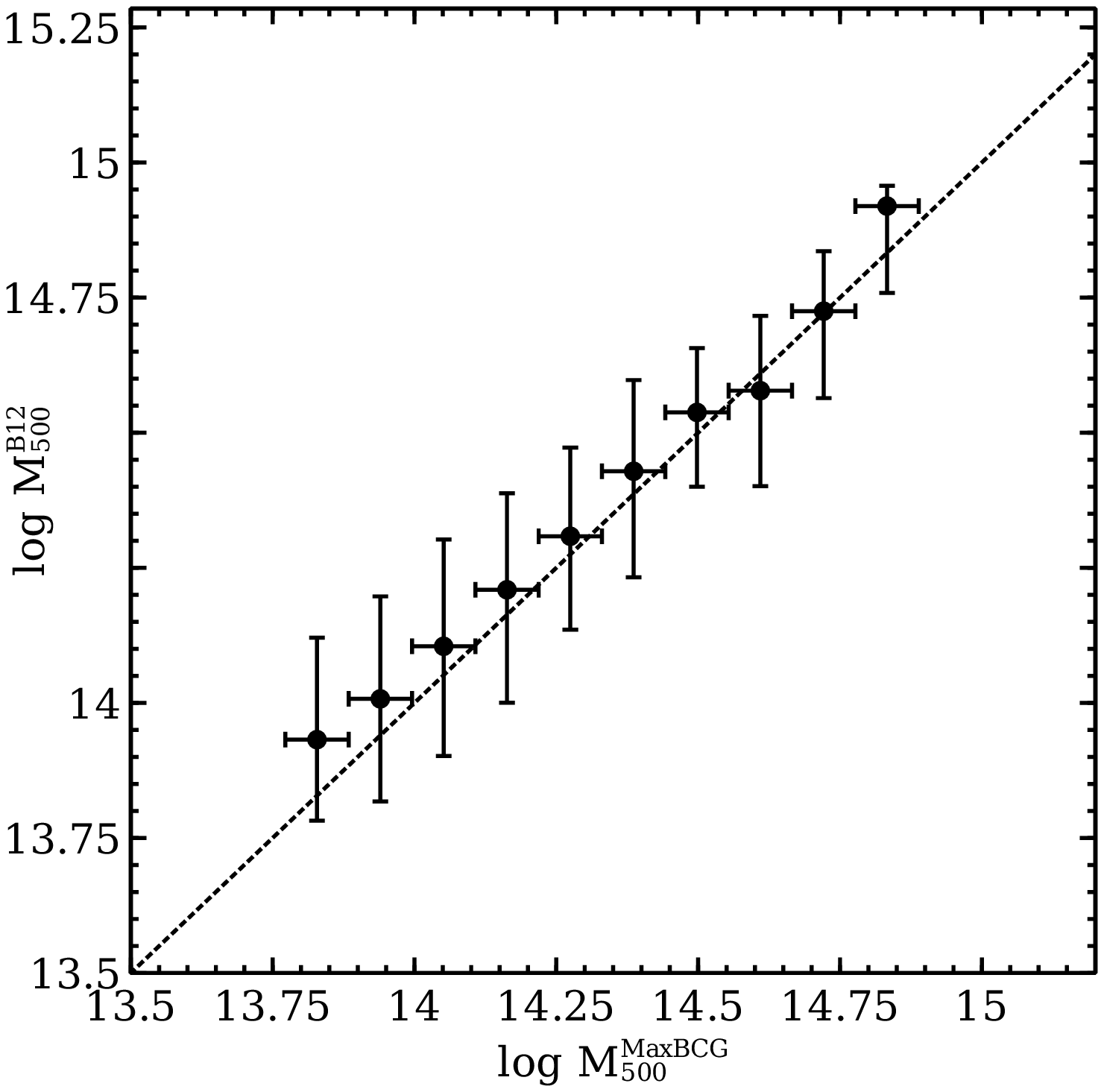}
\caption{Testing the total mass estimates from \citet[][B12]{2012MNRAS.423..104B}. Left-panel: Comparison between the stacked velocity dispersion estimates of the observed LRG group/cluster sample with group/cluster halos extracted from the Millennium Simulation in mass bins from $10^{13.7}-10^{15.0}$ $M_{\odot}$. The dashed green line is the relation from \citet{2008ApJ...672..122E}. Right-panel: A mean comparison with the mass estimates for the overlapping
maxBCG cluster sample from joint X-ray and weak lensing stacking \citep{2009ApJ...699..768R}. The error bars show the 16 and 84 percentiles.}
\label{cap:mass_test}
\end{figure*}

\section{The cluster sample}\label{sec:sample}

In this work we make use of the Luminous Red Galaxy group and cluster sample
described in B12.  The catalogue \footnote{
http://www.ast.cam.ac.uk/ioa/research/cassowary/lrg\_clusters/lrg\_clusters\_dr7.fits
} describes \samplesize \ high-mass groups and clusters with $M_{500}>10^{13.7}$
$M_{\odot}$ within the redshift range $0.15 \leq z \leq 0.4$, and each cluster
is attributed a total mass estimate obtained using a calibration between the
optical richness of bright galaxies ($M_{r}\leq-20.5$) within 1 Mpc, and the
X-ray derived mass for a small subset of \xraysamplesize \ groups and clusters
with X-ray measurements.  Specifically, in their Eq.\ 3, B12 find

\begin{equation}\label{eq:masscalib}
\log \left(\frac{M_{500}}{M_{\odot}}\right) = m \log N_{1 {\rm Mpc}} + b \; ,
\end{equation}

\noindent where the best-fit values are found to be $m=1.4 \pm 0.1$ and
$b=12.3 \pm 0.1$ using the robust fitting algorithm of \citet{2010arXiv1008.4686H}.  The mass threshold of $M_{500}=10^{13.7}$ $M_{\odot}$ corresponds to a richness of $\approx 10$.

The catalogue is shown to be roughly 97 per cent pure and 90 per cent complete above this $10^{13.7}$ $M_{\odot}$ limit (see B12 for more details).

\subsection{Testing the accuracy of the $M_{500}$ estimates}\label{sec:testmasses}

Accurate total mass estimates are required not only because we want to know the total mass of the systems under consideration and to be able to measure stellar mass {\it fractions}, but also because the aperture within which the stellar mass is measured (i.e., $r_{500}$) is determined from the total mass.  Below we test the accuracy of our $M_{500}$ estimates by: (i) deriving stacked velocity dispersion measurements of our groups and clusters and making comparisons with stacked velocity dispersions of simulated groups and clusters extracted from the Millennium Simulation; and (ii) comparing our derived total masses with stacked weak lensing-derived total masses for a large subset of groups and clusters in common between our sample and that of the maxBCG sample.

For the stacked velocity dispersion test, we first query all spectroscopic galaxies within a projected $r_{500}$ for all clusters in our sample and then calculate the corresponding peculiar velocities relative to the LRG using the prescription described in \citet{1999astro.ph..5116H}.  We then create a histogram of velocities by stacking all the cluster fields in a given bin, and fit a model to obtain estimates for the line of sight velocity dispersion, $\sigma$.  The model includes a Gaussian component for the contribution to the velocity dispersion due to the cluster, and a constant component for the contribution from interlopers.

If our total masses are accurate, then we expect that our stacked velocity dispersions should agree with stacked velocity dispersions of groups and clusters of the same mass in (N-body) cosmological simulations.  In particular, we select group/cluster halos with $M_{500} > 10^{13.7} M_\odot$ from the $z=0.32$ snapshot of the Millennium Simulation \citep{2005Natur.435..629S} for analysis.  We extract all central and satellite galaxies with stellar masses $M_{\rm star} > 10^9 M_\odot$ from the corresponding \citet{2007MNRAS.375....2D} semi-analytic galaxy catalog that are within a projected radius of $r_{500}$ of the cluster that have a absolute physical velocity (i.e., peculiar velocity + Hubble flow) $< 2500$ km/s with respect to the cluster center of mass velocity.  The same cut is applied to the peculiar velocities of the observed cluster members.  We estimate the velocity dispersion in precisely the same way as done for the observed groups and clusters (i.e., fitting a Gaussian + constant model).
We note that the derived velocity dispersions are quite insensitive to the stellar mass cut adopted for the simulated galaxies, but they are somewhat sensitive to the velocity cut imposed (for both the simulations and the observations).  However, so long as we adopt the same velocity cut for the simulated and observed galaxies the comparison is a fair one.

Importantly, the values of $M_{500}$ and $r_{500}$ used in our stacking analysis of the simulations correspond to their true values, while for the observed clusters the values of $M_{500}$ and $r_{500}$ are inferred from the observed richness within 1 Mpc.  By comparing the stacked velocity dispersions of the simulated clusters to the observed ones we are therefore testing the accuracy of our total mass estimates for the observed groups and clusters.  If, for example, our observed total mass estimates are on average biased low, then we should expect to see a higher normalisation in the observed $\sigma-M_{500}$ relation compared to the simulated one.

A comparison of the velocity dispersions within $r_{500}$ for the simulations and observations is shown in Fig. \ref{cap:mass_test} (left-panel).  The solid red circles represent the stacked $\sigma-M_{500}$ relation derived from the observed groups and clusters in our sample.  The solid blue circles correspond to the stacked relation using simulated groups and clusters from the Millennium Simulation.  The small black dots represent individual simulated groups and clusters.  The dashed green curve represents the $\sigma_{\rm DM}-M_{500}$ relation derived by \citet{2008ApJ...672..122E}, where $\sigma_{\rm DM}$ is the 1D velocity dispersion of dark matter particles which are generally part of the main halo (i.e.,  not in substructures).  Note that we have converted the $\sigma_{\rm DM}-M_{200}$ relation into a $\sigma_{\rm DM}-M_{500}$ relation assuming a NFW profile and the mass-concentration relation of \citet{2008MNRAS.390L..64D}.

Encouragingly, the observed and simulated stacked relations agree to $\la 10$ per cent accuracy in total mass at fixed $\sigma$.  Thus, our observed total mass estimates are quite accurate on average.  Interestingly, the amplitude and direction of the small offset is consistent with that expected from deviations from hydrostatic equilibrium due to turbulence and non-thermal pressure support in general in the
intra-cluster medium\footnote{Here we remind the reader our total mass estimates are calibrated on a richness-hydrostatic mass relation.} (e.g., \citealt{2007ApJ...655...98N}).  

While the simulated stacked $\sigma-M_{500}$ relation is encouragingly similar to the observed one, indicating a robustly determined average total mass, we note that the scatter in the $\sigma-M_{500}$ relation of {\it individual} simulated clusters is very large ($\sigma_{\rm RMS} \approx 84$ per cent for haloes with $M_{500} \sim 10^{14} M_\odot$).  This should serve as a warning for studies with relatively small samples of clusters and that estimate total masses based on galaxy dynamics.  We will touch upon this again in Section 4.

We also compare the results to the $\sigma_{\rm DM}-M_{500}$ relation found by \citet{2008ApJ...672..122E} and find remarkably good agreement (compare the solid blue and red circles with the dashed green line), adding further confidence to the estimated total masses of our observed groups and clusters.

A further independent check of our mass estimates can be made by comparing our clusters masses with estimates for those systems also found in the
maxBCG sample \citep{2007ApJ...660..239K}. A calibration between $M_{500}$ and optical richness is provided by \citet{2009ApJ...699..768R}, who use stacked weak lensing to constrain the total mass in richness bins.  We compare the maxBCG masses with our masses for a large overlapping subsample of $\sim4500$ systems in Fig. \ref{cap:mass_test} (right-panel) on a cluster-by-cluster basis, and find good agreement within the error bars across the entire range of total masses.  The level of agreement is actually slightly better than expected since, as noted above, non-thermal pressure support is expected to bias X-ray hydrostatic masses within $r_{500}$ low by $\sim10-15$ per cent.  A possible reason for the slightly better-than-expected agreement is that weak lensing itself does not yield an unbiased mass measurement.  Indeed, both \citet{2011ApJ...740...25B} and \citet{2012MNRAS.421.1073B} have found using cosmological simulations that deviations at large radii in the mass profile of clusters from the NFW parametric form adopted in many weak lensing studies leads to a (negative) bias in the reconstructed mass at the level of $\approx 5$ per cent.

\section{Stellar mass estimates}
 
In order to investigate how the stellar mass varies as a function of total mass in our sample, we stack the clusters in four mass bins ranging from $10^{13.7}$ to $10^{15.0}$ $M_{\odot}$. The four bins have approximately equal width in log-space (13.7-14.0, 14.0-14.4, 14.4-14.7, 14.7-15.0) and shown in B12 the scattering of clusters between mass bins (due to errors in the recovered mass) is minor (see their Fig.\ 6).

\subsection{Method 1: Stacking the galaxy catalog}\label{sub:photostack}

For each cluster field we follow the prescription described in B12 and obtain all SDSS photometric and astrometric data for galaxies by running the \texttt{q3c} radial query algorithm \citep{2006ASPC..351..735K} on a locally available SDSS DR7 database. The galaxies are queried within a physical (projected) 5 Mpc aperture and have the following properties:

\begin{equation}
r<21.5,\\
|z-z_{\mathrm{BCG}}| \leq 0.04(1+z_{\mathrm{BCG}}),\\
\sigma_{z}<0.2,
\label{eq:neighcond}
\end{equation}

\noindent where $r$ is the extinction-corrected $r$-band model
magnitude, $z$ is the photometric redshift of neighboring galaxies and
$\sigma_{z}$ is the corresponding photometric redshift
error. For each galaxy we then calculate the K-corrected absolute magnitude (relative to the redshift of the LRG) using the prescription of \citet{2010MNRAS.405.1409C}, and retain only galaxies with $M_{r}\leq-20.5$ to ensure completeness of the sample at all redshifts (see \citealt{2012MNRAS.423..104B}). Estimates of the background density of galaxies for each cluster are obtained by querying random fields at the same redshift of each system.

The stellar mass for each individual galaxy is then estimated using the colour-dependent mass-to-light prescription of \citet{2003ApJS..149..289B}:

\begin{equation}
\mathrm{log_{10}}\left(M/L_{r}\right)=-0.306+1.097\left(g-r\right)-0.1\;,
\label{eq:bdjm2l}
\end{equation}

\noindent where $M/L_{r}$ is the mass-to-light ratio in the SDSS $r$-band, $g-r$ is a rest-frame colour, and the 0.1 factor is to convert from a `diet-Salpeter' to a \citet{2003PASP..115..763C} IMF. For each total mass bin we then stack the fields to obtain an average stellar mass within an aperture of radius equal to the mean $r_{500}$ in each bin, and subtract a characteristic background density estimate (i.e., from stacking random fields).

The simplicity of the \citet{2003ApJS..149..289B} prescription provides an attractive method for comparing our results with those of other authors in a self-consistent way.  However, the method of fitting simple parametric galaxy evolution models to galaxy colours (based on an earlier paper by \citealt{2001ApJ...550..212B}), uses spiral galaxies in the model, and it is clear that our groups and clusters contain predominantly red galaxies (see Table \ref{tab:brightscols}). Therefore, we must test how accurately the \citet{2003ApJS..149..289B} prescription recovers the stellar masses of these red galaxies by comparing the mass estimates on a galaxy-by-galaxy basis.

\begin{figure}
\includegraphics[width=\columnwidth]{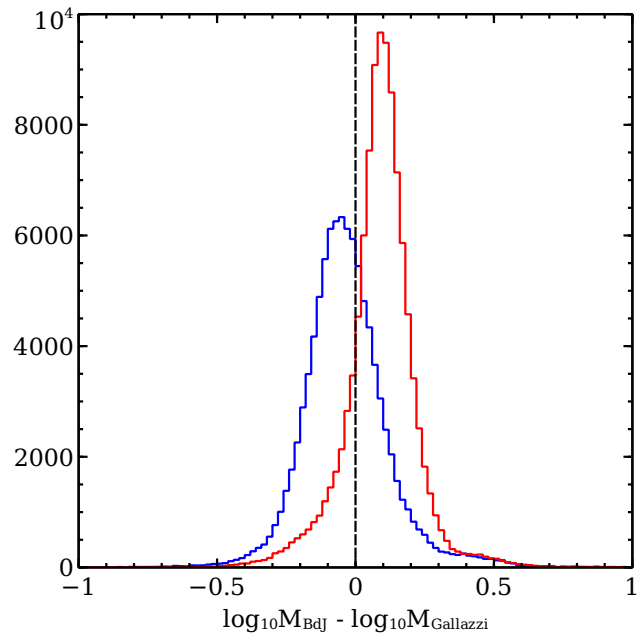}
\caption{A comparison of the stellar mass estimates of galaxies from the \citet{2005MNRAS.362...41G} catalogue, compared with the estimates obtained using the prescription of \citet{2003ApJS..149..289B}. The plot shows the difference between the estimates for red galaxies (red) and blue galaxies (blue).}
\label{cap:galmatch}
\end{figure}

\citet{2005MNRAS.362...41G} provide more accurate stellar mass estimates for $\sim180\,000$ galaxies from SDSS DR2 by providing direct model fits to the $H\delta_{\alpha}$ and $D4000$ spectral absorption features. We can therefore test the accuracy of the \citet{2003ApJS..149..289B} prescription by calculating the corresponding stellar mass for the \citeauthor{2005MNRAS.362...41G} galaxies by using eqn.~\ref{eq:bdjm2l}. The comparison is shown in Fig. \ref{cap:galmatch}. Splitting the \citeauthor{2005MNRAS.362...41G} galaxy sample into red and blue galaxies (using the colour cut below in eqn.~\ref{eq:redseq}) we find that on average the stellar mass of the red galaxies (i.e. the major component in our groups) is overestimated by $\sim$0.1 dex. The opposite behaviour is observed for blue galaxies, whose stellar masses are underestimated by $\sim$0.05 dex. These small corrections are applied when calculating the stellar mass fractions of groups and clusters in our cluster sample.

\subsubsection{Correction for stellar mass in galaxies fainter than $M_r = -20.5$}\label{sub:missflux}

\begin{figure}
\includegraphics[width=\columnwidth]{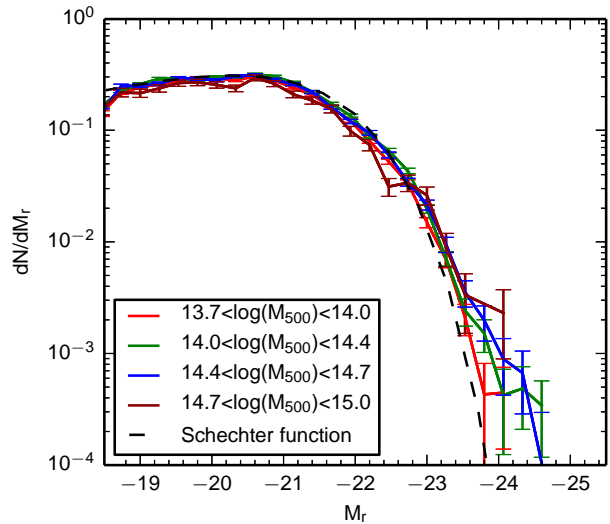}
\caption{The stacked luminosity distributions of group member galaxies within $r_{500}$ in the group/cluster sample. The sample is split into bins of total mass and restricted to redshifts of $z<0.2$ to allow for probing of the faint end slope. The shape (particularly that of the faint end slope) does not change significantly with total mass. The best fit Schechter function (Equation \ref{eq:schecfunc}) is overplotted as a black dashed line.}
\label{cap:grouplfs}
\end{figure}

To correct for the missing flux corresponding to faint cluster member
galaxies with $M_{r}>-20.5$, we require information about the cluster galaxy
luminosity distribution and in particular, the faint-end slope of the
distribution.  We therefore produce a stacked background-subtracted
luminosity distribution as a function of mass (Fig.  \ref{cap:grouplfs}),
 for all cluster members excluding the central LRG.  We
use only lower redshift groups with $z<0.2$ to ensure we are complete to a
fainter magnitude limit ($M_{r}=-18.5$).

We find no significant change in the shape of the luminosity distribution with increasing cluster mass, and therefore we fit a \citet{1976ApJ...203..297S} function (with arbitrary normalization) of the following form to a single large mass bin:

\begin{eqnarray}
\Phi\left ( M \right )& \propto & 10^{0.4\left ( \alpha +1 \right )\left ( M^{\star}-M \right )} \nonumber \\
& & \times \exp \left ( -10^{0.4\left ( M^{\star}-M \right )} \right),
\label{eq:schecfunc}
\end{eqnarray}

\noindent with $M^{\star}=-21.41 \pm 0.07$, $\alpha=-0.68 \pm 0.04$.  Although this functional form does not fit the bright end well\footnote{Previous studies have found that the presence of central bright galaxies leads to an excess relative to the Schechter function (e.g., \citealt{2005A&A...433..415P,2011MNRAS.412..947B}).}, the fit to intermediate and lower luminosity galaxies is quite good.  We integrate the best-fit Schechter function down to zero and find a missing flux percentage of 20 per cent (i.e., 20 per cent of the flux of the total flux obtained by integrating over all luminosities comes from galaxies finater than -20.5).  The missing flux percentage is found be insensitive to the integration lower limit, e.g., the correction factor is found to differ by only 1 per cent if we adopt a lower luminosity cut off of -18.5 (i.e., the limit to which we are complete in Fig.~\ref{cap:grouplfs}).

The form of the luminosity distribution in Fig. \ref{cap:grouplfs} is
comparable to many studies of groups and clusters in the literature.  Our
value of $M^{\star}$ is in excellent agreement with those obtained by
\citet{2009ApJ...690.1158C}, \citet{2009ApJ...695..900Y} and
\citet{2011MNRAS.414.2771D} in a variety of studies on SDSS groups and
NoSOCS clusters.  Although shallower, our faint-end slope is also consistent
with that of \citet{2009ApJ...690.1158C} who find $\alpha=-0.84 \pm 0.32$.

Note, however, \citet{2005ApJ...633..122H} and \citet{2011MNRAS.414.2771D} find that $\alpha$ is consistent with $-1.00$, and that the shape of the luminosity distribution changes with cluster richness. This discrepancy is plausibly explained by the that these authors obtained the stacked luminosity distribution within different apertures \citep[see argument by][]{2011MNRAS.414.2771D}.

\subsection{Method 2: Stacking SDSS images}\label{sub:imagestack}

We also calculate stellar masses by stacking the light in SDSS images. This is
independent of the galaxy stacking method described in Section
\ref{sub:photostack} and has the added advantage of not requiring a correction
for the missing flux (Section \ref{sub:missflux}). This calculation will also
implicitly include any ICL component present in our clusters.

The stacking prescription is summarised as follows, with a full description of
the systematic effects to be presented in a forthcoming paper. The key steps
of the procedure are:

\vskip0.1in

\noindent i) {\bf Image retrieval.}
For each LRG in our catalog, we retrieve the reduced, sky-subtracted and flux
calibrated $r$ band images covering the LRG and the area around it from the
SDSS-III archive.

\vskip0.1in

\noindent ii) {\bf Sky subtraction.}
All the SDSS-III images have their background subtracted.   Although SDSS-III
sky subtraction algorithm was improved relative to SDSS-I/II and performs very
well in general \citep{sdss3}, when studying very extended structures such as ICL or total light of the galaxy groups the SDSS background estimates may be biased, because
the SDSS-III sky determination algorithm uses the sky near the groups to build
the global polynomial sky model across multiple fields. And although the SDSS
pipeline masks all the sources when constructing the global sky model, it is
possible that the resulting sky is still overestimated due to the presence of
the low surface brightness ICL and undetected galaxy population in the group/cluster.
Therefore, instead of using global model of the background provided
by SDSS, we estimate the sky level at the location of the system based on
the sky in the fields in the same stripe, but located far enough
from the center of the system (more than 15\arcmin\ or $\sim $ 3\,Mpc at z=0.2). Although
this procedure makes our background estimates slightly more noisy than the SDSS
ones, they are not expected to be biased by the ICL or
undetected galaxies present in the clusters/groups.  

\vskip0.1in

\noindent iii) {\bf Mosaicing.}
The sky-subtracted images from the previous step are rebinned to a fixed
physical pixel size of $\sim$ 1.3\,kpc (which corresponds to the SDSS pixel of
0.\arcsec396 at redshift of 0.2) and large (7000x7000 pix$^2$) mosaics centered
on
every LRG in our catalog are created.

\vskip0.1in

\noindent iv) {\bf Masking.}
Bright Milky Way stars in the images are obviously an undesirable contaminant,
so we mask all the stars brighter than $r<$21.5 by circular apertures of
different sizes. The size of the apertures have been derived from the
magnitudes of the stars assuming power-law wings \citep{sdss3} and designed to
leave the PSF wings fainter than $\sim$30\,mag/sq.arcsec. The reason for masking
only stars that are brighter than 21.5 is that near this threshold the
star/galaxy separation algorithms of SDSS start to break down \citep{sdssedr}, so
we could artificially mask faint galaxies if we were to go fainter. Additionally, we
mask bright UCAC \citep{ucac} stars and extended RC3 galaxies \citep{rc3} which
are close enough to our fields, by cross-matching with those catalogs.
And finally we use the SDSS masks (fpM files) which identify certain defects
such as satellite trails and defects arising because of bright stars to further
clean up the images. This ensures that the flux left in our images is mostly
coming from the galactic background/foreground and from the groups
that we study (and possibly from a small number of faint stars).

\vskip0.1in

\noindent v) {\bf Stacking.}
Before stacking, the background-subtracted, rebinned, and masked mosaics of galaxy groups are corrected for different luminosity distances of the central
LRGs.  We then average images in 4 bins of halo masses $M_{500}$ (the bins are
given in Tab.~\ref{tab:fitres} and \ref{tab:brightscols} ). The pixels that were
masked in the previous step were skipped by the averaging procedure.

\begin{figure*}
\includegraphics[width=\textwidth]{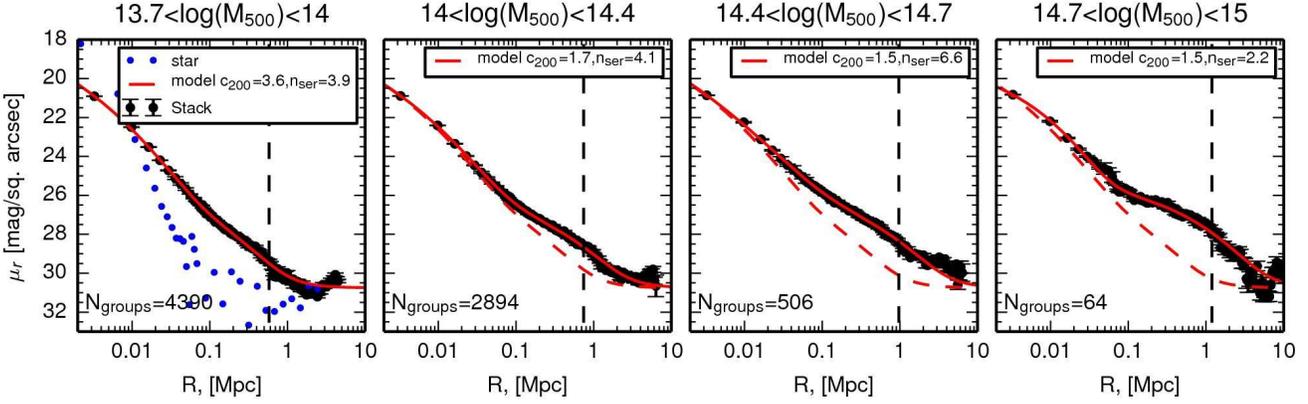}
\caption{The stacked $r$-band surface brightness as function of radius for all
four mass bins in our sample (black points). We stack using the prescription
described in Section \ref{sub:imagestack}. The solid red line shows the best
fitting Sersic$+$NFW model in each mass bin, and the dashed red line shows the
lowest mass bin model repeated. The blue points show a stacked set of stars
with $r\sim 17.5$ from SDSS for comparison. The dashed lines indicates the
r$_{500}$ corresponding to the average halo mass of the bins.}
\label{cap:stacklight}
\end{figure*}

\vskip0.1in

\noindent vi) {\bf Star check.}
As a check of correctness of the stacking methodology we have also tested our code on a randomly distributed set of $\sim$1000 fields centered on stars with a narrow range of magnitudes $17.4<r<17.5$ to verify that we correctly recover the average magnitude of the star from the stack (the profile from the stack are shown as blue circles in Fig.~\ref{cap:stacklight}).

\vskip0.1in

\noindent vii) {\bf Surface brightness profiles.}
The stacked 2D image for each mass bin are azimuthally-averaged to produce
the 1-D surface brightness profiles. The resulting radial surface brightness
profiles together with their fits in the $r$ band are shown in Fig.
\ref{cap:stacklight}.  The figure shows a very noticeable trend with increasing halo mass:  While at low halo masses the surface brightness tends to be dominated by the central LRG, at higher masses, we see the appearance of a large ``bump'' at $\lesssim r_{500}$ due to the increasing importance of the satellite galaxy population.

\vskip0.1in

\noindent viii) {\bf Parametric modelling.}
The resulting surface brightness profiles have the
contribution from the central LRGs, ICL, and from all the satellite galaxies in our systems but also light from all background and foreground galaxies along the line of sight. In order to describe these profiles we choose a composite parametric model consisting of the sum of a Sersic profile (to represent the central LRG), an NFW profile (to represent the ICL+satellite galaxy population) and a flat background (to represent the uncorrelated foreground/background galaxy population).  This model is fitted to the total light profiles in each of the 4 halo mass bins and reproduces the observed profiles very well.  The background flux levels in 4 different bins and the
background determined from the stack of stars are used to estimate the average
background and its variance.  We then integrate the total flux for each
mass bin to the $r_{500}$. In order to estimate the uncertainty on the total
fluxes, we perturbed the 1-D profiles by their uncertainties as well as
perturbed the background level by the previously estimated variance. The 
resulting measurement of total absolute magnitudes together with Sersic indices 
and NFW concentration parameters for 4 halo mass bins are given in Table~\ref{tab:fitres}.

\begin{table}
\caption{Summary of the fits to stacked r-band surface brightness profiles in $\log_{10}(M_{500}/M_\odot)$ bins.}
\begin{center}
\begin{tabular} {rrrrr} 
\hline 
$\log M_{500}$ & 13.7-14.0 & 14.0-14.4 & 14.4-14.7 & 14.7-15.0\\
\hline 
$\left<M_r\right>^\star$ & -24.29$\pm$0.11 & -25.2$\pm$0.07 & -26.05$\pm$ 0.06 &
-26.83$\pm$0.06\\
$c_{200}$  & -           & 1.7 $\pm$ 0.7 & 1.5 $\pm$ 0.8 & 1.5$\pm$0.2 \\
$n_{sersic}$& 4.8$\pm$ 1.1 & 4.1 $\pm$ 0.7 & 6.6 $\pm$ 1.4 & 2.2$\pm$0.1\\
\hline 
\end{tabular}
\end{center}
\medskip

$\star$ $\left<M_r\right>$  refers to the total r-band 
light from the Sersic and NFW fit within $r_{500}$. 
$c_{200}$ is the fitted concentration of the NFW 
component of the  profile, 
while $n_{sersic}$ is the Sersic index of the Sersic component. 
The numbers refer to the mass bins ranging from $10^{13.7}$
to $10^{15.0}$ $M_{\odot}$ described in Section \ref{sec:sample}. 
For lowest mass bin we have failed to measure concentration
with reasonable accuracy.\label{tab:fitres}
\end{table}

Finally, we use the r-band absolute magnitudes within $r_{500}$ to  calculate
the stellar mass in each halo mass bin using the Bell et al.\ prescription in
Section \ref{sub:photostack} using the mean $g-r$ colour\footnote{We decided
against
using the g-band stack and to use the g-r from the galaxy population since the 
g-band stack had significantly lower signal to noise}
for galaxies in that halo mass bin (see Table 2), and apply the Gallazzi
correction discussed in Section \ref{sub:relations}
We emphasize that in the above analysis the objective is to measure the {\it
total} stellar mass (i.e., from the BCG+ICL+satellites).  We therefore avoid
the complication of having to separate the ICL from the BCG and satellite
population in this study.

\subsection{Deprojection correction}\label{sub:deproj}

The aim of this work is to provide estimates of the stellar mass within a
sphere of radius $r_{500}$ (where $r_{500}$ is the mean within each total
mass bin).  However, what we have measured thus far is a two-dimensional
projected stellar mass estimate (i.e., within a cylinder).  We calculate the
deprojection correction factor as:

\begin{equation}
\frac{M_{3d}}{M_{2d}} = \frac{\int_{0}^{r_{500}}\rho\left ( r \right )4\pi r^{2}\mathrm{d}r}{\int_{0}^{r_{500}}\Sigma\left ( R \right )2\pi R\mathrm{d}R}
\label{eq:nfwcorr}
\end{equation}

\noindent where $\rho$ is the three-dimensional stellar mass density profile
\citep{1997ApJ...490..493N} and $\Sigma$ is the projected stellar surface
mass density.  For simplicity, we assume the stellar 3D mass density profile 
follows a NFW distribution \citep{1997ApJ...490..493N}.  The corresponding
surface mass density profile is given in \citet{1996A&A...313..697B}, which
provides a good fit to the stacked surface brightness profiles as we have shown
above. The ratio in eqn.~\ref{eq:nfwcorr} will depends only weakly on the
concentration of the NFW profile, which we measure from stacked light profile
(Fig. \ref{cap:stacklight}), or from fitting the galaxy density profiles
\citep{2012MNRAS.423..104B} ( $c_{200} \approx 2$ ).  Given this concentration,
the deprojection factor is found to be $\approx$\deprojection. 

\subsection{The bias introduced by adopting a single $M/L$ ratio}

To facilitate comparisons to previous work we have experimented with calculating the stellar masses of our groups and clusters by applying a single $M/L$ ratio uniformly to all galaxies.  Specifically, instead of calculating the stellar mass of each individual galaxy, we obtain the total background-subtracted luminosity within the mean $r_{500}$ in a given total mass bin and then apply eqn.~\ref{eq:bdjm2l} using the corresponding \emph{mean} luminosity-weighted colour in each bin. 

In a given total mass bin we find that stellar mass derived from our preferred method (i.e., which allows for galaxy population variation) is systematically larger than that derived assuming a fixed $M/L$, but only by $\approx$ 5 per cent on average.  Thus, the bias introduced by adopting a single $M/L$ is relatively small in comparison to other potential systematic uncertainties, such as choice of stellar IMF.

\begin{figure*}
(a)
\includegraphics[width=0.47\textwidth]{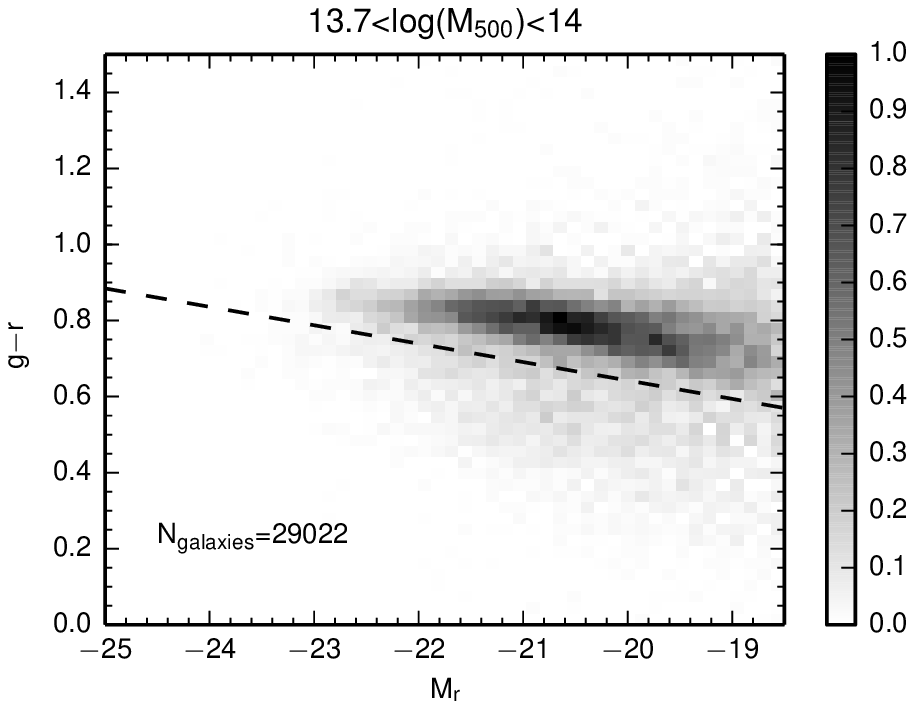}
(b)
\includegraphics[width=0.47\textwidth]{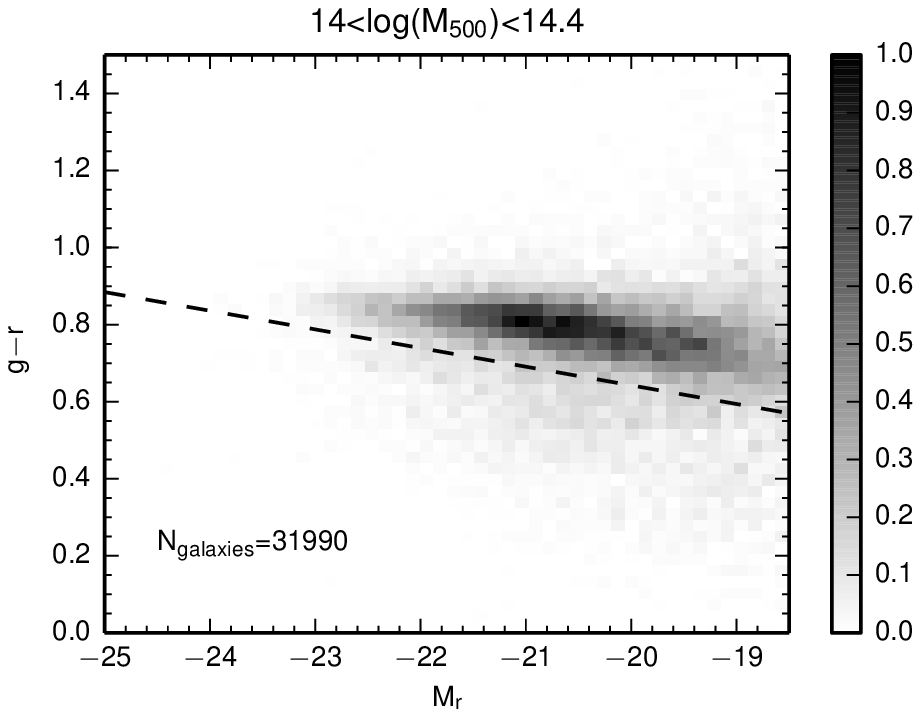}

(c)
\includegraphics[width=0.47\textwidth]{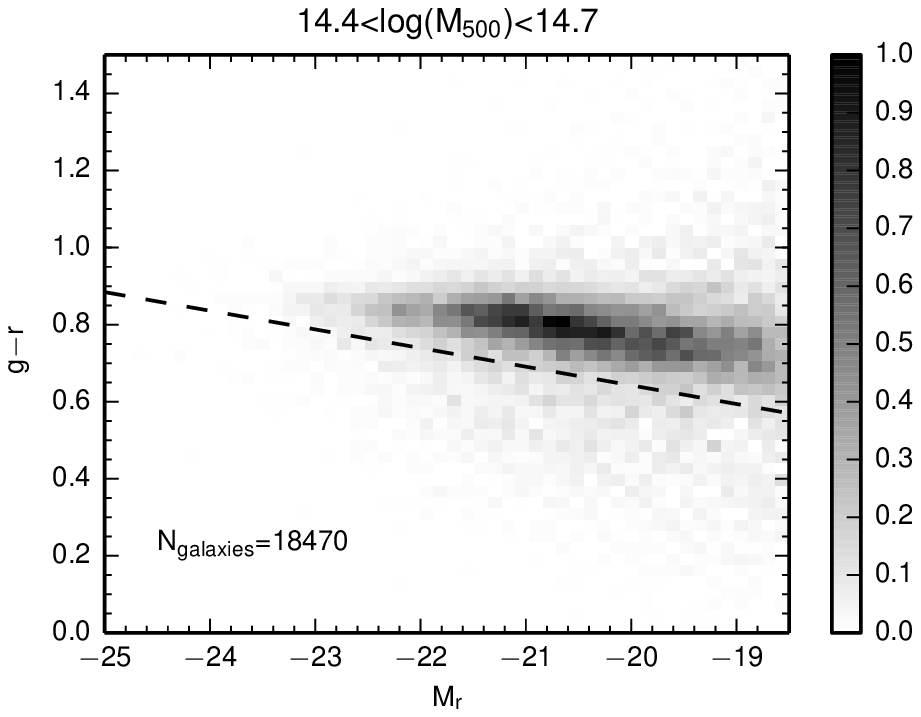}
(d)
\includegraphics[width=0.47\textwidth]{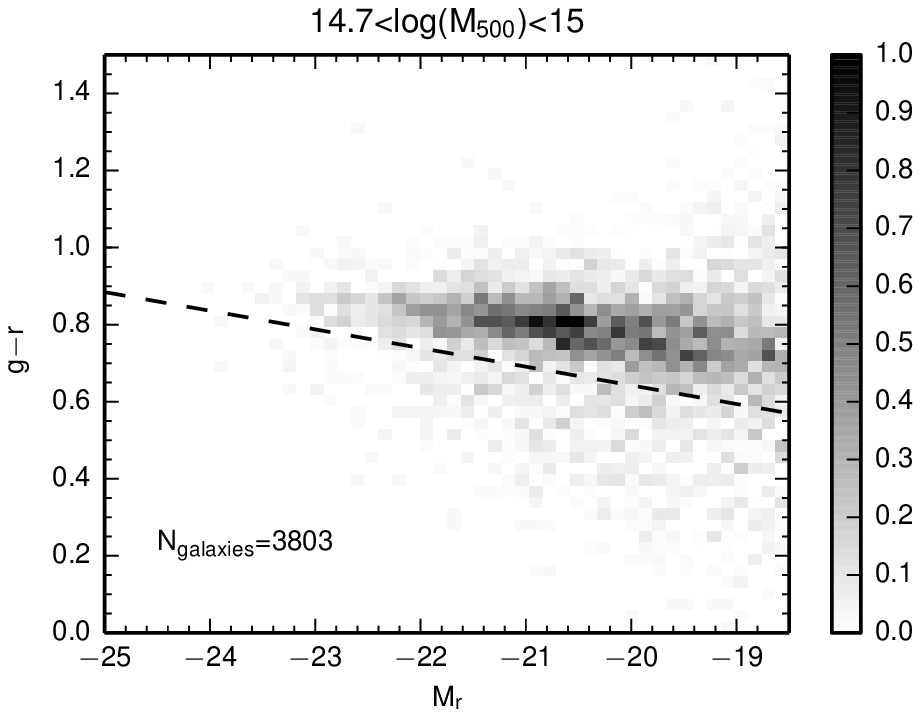}

\caption{The background subtracted stacked colour-magnitude diagrams of group and cluster galaxies within $r_{500}$ as a function of total mass. Panels (a) to (d) show bins of increasing total mass ($\log_{10}M_{500}$ = 13.7-14.0, 14.0-14.4, 14.4-14.7, 14.7-15.0). The dotted line represents our arbitrary split between `red' and `blue' galaxies. The plots have been normalised to compare the shape of the red sequence, which is shown not to change significantly with total mass.}
\label{cap:colourmags}
\end{figure*}

In addition to quantifying the mean bias, we can also examine how the bias changes with system mass.  It is known, for example, that the `quenched' fraction of galaxies increases with system mass.  Thus, if one applies a uniform $M/L$ to all galaxies this will introduce a bias in the slope of the derived $M_{\rm star}-M_{500}$ and $f_{\rm star}-M_{500}$ relations.  To investigate this point further, we examine the mixture of galaxies contributing to these clusters over the range of total masses in terms of mean red/blue fraction, luminosity-weighted colour, magnitude and mass-to-light ratio. We first create a stacked background-subtracted colour-magnitude diagram for each bin (Fig. \ref{cap:colourmags}), and provide a simple red-sequence definition where red galaxies are defined by the region:

\begin{equation}
g-r > -0.048\left(M_{r}\right) - 0.326 \;.
\label{eq:redseq}
\end{equation}

This functional form approximately separates the red sequence from the blue star-forming ``cloud'' in histograms of $g-r$ in magnitude bins.  Our results are not at all sensitive to the particular cut adopted.

Although the cluster red sequence does not appear to change shape significantly across the range of total masses in Fig. \ref{cap:colourmags}, we find a slight change in the mean populations of galaxies in the colour-magnitude plane, which are summarised in Table \ref{tab:brightscols}.  Despite no detectable change in the mean brightness of satellite galaxies with total mass, we see a small change in the red fraction and the $r$-band mass-to-light ratio of around 8 per cent over the mass range. We use the colour scalings of \citet{2003ApJS..149..289B} to infer the corresponding mass-to-light ratio in the $K$-band. We find no detectable change in this ratio ($\approx 0.7$) over the range in total masses we have explored.  We also look at the corresponding change in the Johnson $I$-band and find a modest variation over the range of total masses (see Table \ref{tab:brightscols}).

Therefore, over the range of total masses and redshifts considered here, the adoption of a fixed $M/L$ does not appear to significantly bias the derived stellar masses, so long as the adopted $M/L$ corresponds to the approximately the mean value.  However, the bias may reasonably be expected to become more sigificant when pushing to lower halo masses and/or higher redshifts than considered here, due to the increasing fraction of active (star forming) galaxies in these regimes.

\begin{table}
\caption{Summary of the stellar properties of the groups in $\log_{10}(M_{500}/M_\odot)$ bins. }
\medskip
\begin{center}
\begin{tabular} {r r r r r} 
\hline 
$\log_{10}M_{500}$ & 13.7-14.0 & 14.0-14.4 & 14.4-14.7 & 14.7-15.0\\
\hline 
$\left<M_{r}\right>$ & -21.3 & -21.3 & -21.3 & -21.3 \\ 
$\left<g-r\right>^{\star}$ & 0.77 & 0.78 & 0.79 & 0.80 \\ 
$\left<\mathrm{Red \ fraction}\right>$ & 0.79 & 0.81 & 0.84  & 0.85 \\ 
$f_{\mathrm{mass}}$ & 1.18 & 1.19 & 1.20  & 1.20 \\ 
$\left<M/L_{r}\right>$ & 2.75 & 2.81 & 2.87 & 2.94  \\ 
$\left<M/L_{i}\right>$ & 2.20 & 2.24 & 2.29 & 2.33  \\ 
$\left<M/L_{I}\right>$ & 2.01 & 2.05 & 2.09 & 2.13  \\ 
$\left<M/L_{K}\right>$ & 0.70 & 0.70 & 0.70 & 0.70  \\ 

\hline 
\end{tabular}
\end{center}

\medskip $^{\star}$refers to luminosity-weighted mean colour, and
$\left<M_{r}\right>$ refers to the mean galaxy luminosity above
$M_{r}=-20.5$.  The numbers refer to the mass bins ranging from $10^{13.7}$
to $10^{15.0}$ $M_{\odot}$ described in Section \ref{sec:sample}.  The
mass-to-light ratios are calculated from the mean colours using Equation
\ref{eq:bdjm2l}.  In order to account for the Gallazzi correction (Section
\ref{sub:photostack}), the mass-to-light ratios need to be multiplied by
$f_{\mathrm{mass}}$ from Equation \ref{eq:bellcorr}.

\label{tab:brightscols}
\end{table}

\begin{figure*}
\includegraphics[width=0.45\textwidth]{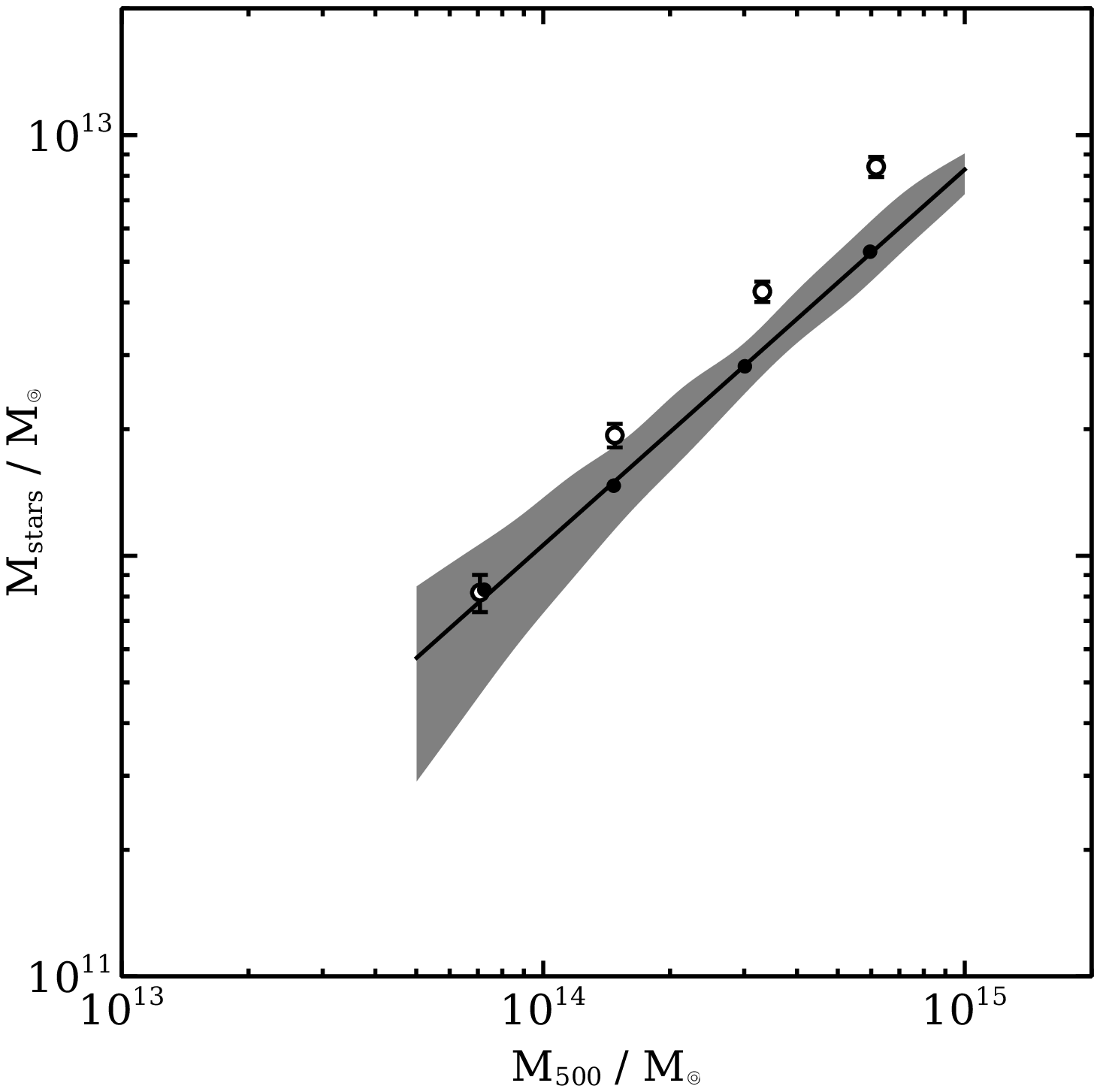}
\hspace{5mm}
\includegraphics[width=0.45\textwidth]{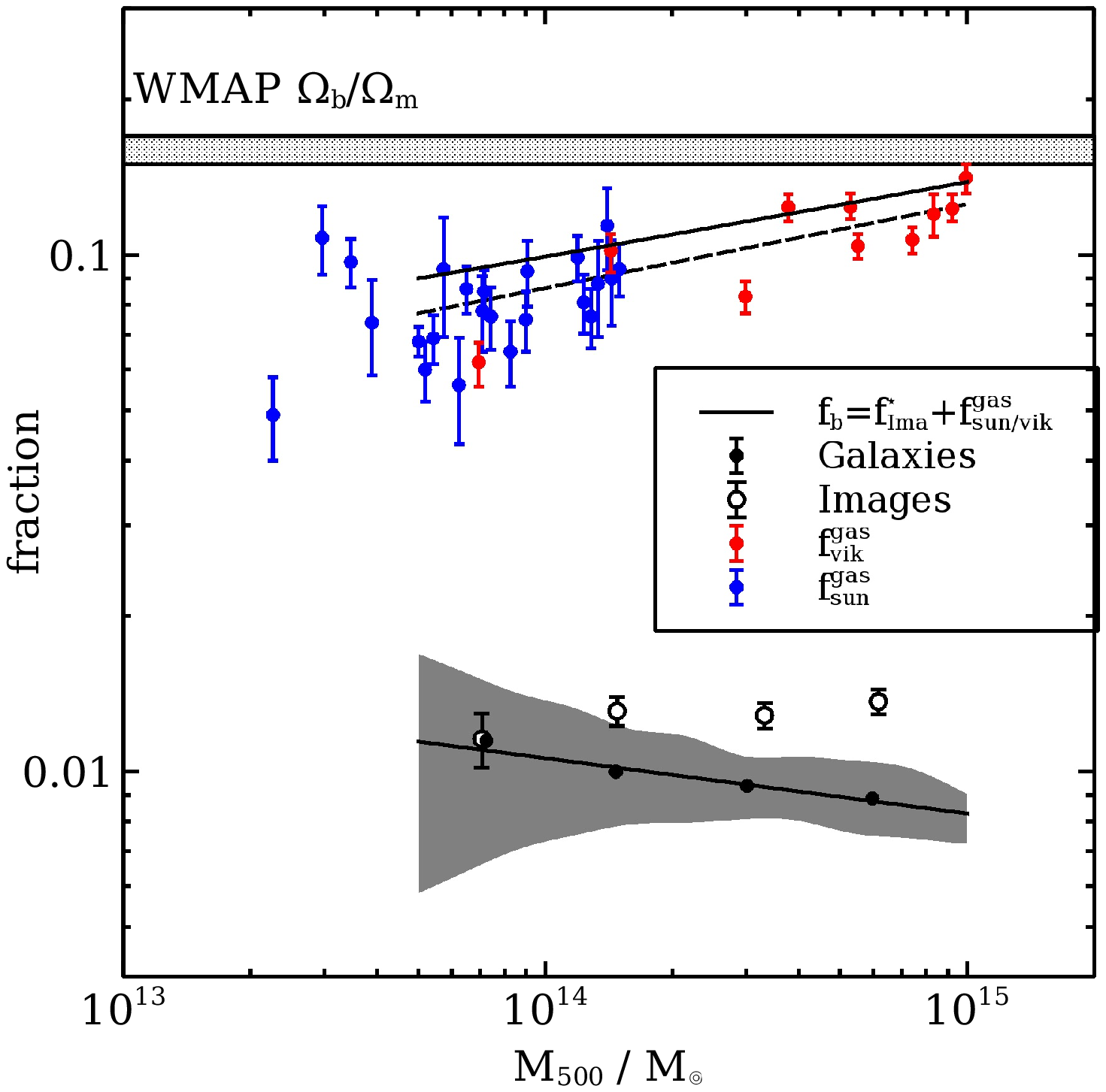}
\caption{The stacked relationship between stellar mass and total mass in a large
sample of groups/clusters split into four bins of total mass. Left-hand panel:
The solid points represent the background-subtracted stellar mass calculated
from the SDSS photometric catalogue, and the empty circles represent the result from
summing the light from the SDSS-III images. The grey region represents an
estimate of the $1\sigma$ error bar obtained by the projection of the error
ellipses in Fig. \ref{cap:scatter} onto the vertical axis. Right-hand panel:
This plot shows that the stellar mass fraction does not depend strongly on total
mass. We overplot measurements of the gas fraction in groups and clusters from
\citet[][blue points]{2009ApJ...693.1142S} and \citet[][red
points]{2006ApJ...640..691V}, as well as a fit to gas mass fraction as a
function of total mass (dotted line). An estimate for the baryon fraction
(solid-line) is provided by summing the contribution from gas and stars
(from the SDSS Images stack). The WMAP 9-yr baryon fraction \citep{2013ApJS..208...19H} is shown for comparison.
}
\label{cap:ourresult}
\end{figure*}

\section{Results}

\subsection{The $M_{\rm star}-M_{500}$ and $f_{\rm star}-M_{500}$ relations}\label{sub:relations}

In order to assess how the stellar mass fraction varies with total mass we
provide a stacked result in four mass bins over the mass range
$M_{500}>10^{13.7}$ $M_{\odot}$ in Fig.  \ref{cap:ourresult}.  The solid
points show the result from the stacked galaxy catalog and the empty circles show
the result from the stacked images.  The following corrections have been
applied.  For both the galaxy catalog and image stacking results we modify
the stellar masses implied by the prescription of
\citet{2003ApJS..149..289B} by a factor

\begin{equation}
f_{\mathrm{mass}}=10^{0.1}f_{\mathrm{red}}+10^{-0.05}\left(1-f_{\mathrm{red}}\right)
\label{eq:bellcorr}
\end{equation}

\noindent so that they agree with the more accurate spectroscopic estimates of \citet{2005MNRAS.362...41G} (see Section \ref{sub:photostack}). The quantity $f_{\mathrm{red}}$ corresponds to the cluster galaxy red fraction which varies as a function of total mass (see Table \ref{tab:brightscols}). For the galaxy catalog method, we then increase the stellar mass by 20 per cent to correct for the mass in galaxies fainter than $M_r = -20.5$ (see Section \ref{sub:missflux}).  Finally, we apply the 30 per cent deprojection correction (see Section \ref{sub:deproj}) to both methods.

As both our estimate of the stellar mass and the total mass depend on the properties of the galaxies (our total mass is estimated using a mass-richness relation), there is undoubtedly a degree of covariance between both axes in Fig. \ref{cap:ourresult}.  We have modelled the degree of covariance and scatter between total mass bins in Appendix \ref{sec:app} and find that, while certainly present, it does not drive our main results.  The shaded regions in Fig. \ref{cap:ourresult} represent an estimate of the $1\sigma$ error bar obtained by the projection of the error ellipses in Fig. \ref{cap:scatter} onto the vertical axis. 

We fit a power law relation of the following form to the galaxy catalogue stacking results (solid points):

\begin{equation}\label{eq:mstellarhalo}
\log \left(\frac{M_{\rm star}}{M_{\odot}}\right) = a \log \left(\frac{M_{500}}{3\times10^{14} M_{\odot}}\right) + b \; ,
\end{equation}

\noindent with best-fit parameters $a=0.89\pm0.14$ and $b=12.44\pm0.03$.

The implied $f_{\rm star}-M_{500}$ relation is therefore given by

\begin{equation}\label{eq:fstellarhalo}
\log f_{\rm star} = \alpha \log \left(\frac{M_{500}}{3\times10^{14} M_{\odot}}\right) + \beta \; ,
\end{equation}

\noindent where $\alpha=-0.11\pm0.14$ and $\beta=-2.04\pm0.03$.

For the image stacking results (emtpy circles) we find $a=1.05\pm 0.05 $ and
$b=12.60\pm0.01$ for the $M_{\rm star}-M_{500}$ relation and $\alpha=0.05\pm
0.05$ and $\beta=-1.88\pm 0.01$ for the $f_{\rm star}-M_{500}$ relation.

Interestingly, we find that the image stacking method yields stellar masses
that are $\sim$20-40 per cent larger than those inferred from the galaxy
catalog method (excluding the lowest mass point with the largest errorbars). 
Since we have taken care to correct for the contribution
from faint galaxies in the galaxy catalog method (in Section
\ref{sub:missflux}), it is tempting to attribute the larger mass inferred
from the image stacking method to an ICL component.  We point out, however,
that some of the additional mass could also be associated with the outskirts
of galaxies in the galaxy stacking method (i.e., if the SDSS `model'
magnitudes systematically underestimate the true luminosity of galaxy). 
Furthermore, we have assumed that the ICL has the same colour (and hence
mass-to-light ratio) as the galaxies.  A more detailed exploration of ICL of
our group and cluster sample will be presented in a forthcoming paper.

We overplot measurements of the gas mass fraction from from \citet{2009ApJ...693.1142S} and \citet{2006ApJ...640..691V} in Fig. \ref{cap:ourresult} (right panel). We also provide a fit and an estimate of the total baryon fraction ($f_{b}$ - the sum of the contribution from gas and stars), which is shown to be a strongly varying function of total mass. We also plot the WMAP 9-yr baryon fraction \citep{2013ApJS..208...19H} for comparison and note that $f_{b}$ is below the WMAP value for all masses and significantly so for galaxy groups.

Previous studies have also concluded that the baryon fraction of groups and clusters lies below the universal baryon fraction (e.g., \citealt{2003MNRAS.344L..13E,2007MNRAS.377.1457M}.  There are at least three (non-mutually exclusive) possible explanations for this behaviour:

\vskip0.1in
\noindent i) The gas mass and/or stellar mass fractions are underestimated.  
It is difficult to see how the gas mass fractions could be significantly 
underestimated, there are some indications to the contrary in fact 
(due to gas clumping, e.g., \citealt{2011Sci...331.1576S}).  As we discuss later in the paper, 
however, there are large relatively large 
systematic uncertainties in the stellar mass fractions, due to uncertainties in stellar population modelling and the IMF.  
Adoption of a Salpeter IMF (as opposed to a Chabrier IMF), for example, would bring the total observed baryon 
fraction within 2-sigma of the WMAP 9-yr results, for the most massive clusters 
at least.  For lower mass groups, which have significantly lower gas mass fractions than 
massive clusters, we conclude the observed baryon fractions cannot be reconciled with the universal baryon
by appealing to known systematic uncertainties in the stellar mass-to-light 
ratio.

\vskip0.1in

\noindent ii) The maximum-likelihood universal baryon fraction from WMAP is biased high (as proposed by, e.g., \citealt{2007MNRAS.377.1457M}).  This would reconcile the difference at the highest masses but does not explain the large deficit at the group scale.

\vskip0.1in

\noindent iii) As discussed in Section 1, AGN feedback can lower the baryon fractions of 
groups and clusters within $r_{500}$, by ejecting gas from their progenitors.  This is seen in full cosmological hydrodynamical simulations
that include efficient AGN feedback \citep{2010MNRAS.406..822M,2011MNRAS.412.1965M}.

All three of these scenarios may play some role in reconciling the baryon deficit in groups and clusters with respect to the
universal baryon fraction.

\begin{figure}
\includegraphics[width=\columnwidth]{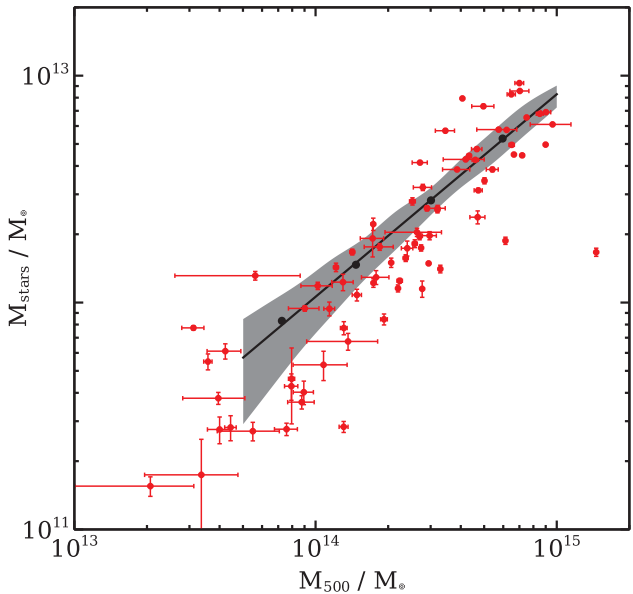}
\caption{The stellar mass vs total mass for the X-ray calibration sample (red points). Overplotted is the galaxy catalogue stacking result (black points) and best fit for the groups/clusters and the grey region represents the $1\sigma$ error bar. We see reasonable agreement over the full range of masses.}
\label{cap:xraycomp}
\end{figure}

\subsubsection{The X-ray calibration sample}

It is of interest to compare our main stacking results with estimates of stellar mass-total mass relation from individual groups and clusters in our richness-mass calibration sample from B12 (i.e., those groups/clusters with X-ray temperature measurements).  The X-ray sample covers a large redshift range from $0.01 \leq z \leq 0.4$, however the stellar mass fractions of clusters (of fixed mass) are not expected to involve much over this range \citep[e.g.][]{2012ApJ...745L...3L}.  The stellar masses are calculated in an identical way to that described above and the total masses are derived from the X-ray temperature - total mass relation by \citet{2006ApJ...640..691V}.  The error bars on the halo mass only take into account the reported uncertainties on the X-ray temperature (i.e., assume a perfect correlation between $M_{500}$ and temperature), while the errors on the stellar mass are statistical only and do not take into account errors in the aperture $r_{500}$ due to uncertainties in $M_{500}$.

A complication arises from the fact that many of our low-temperature groups
are at very low redshift\footnote{This is due to the difficulty in obtaining
robust temperature measurements of low-mass groups at high redshift with the
current generation of X-ray telescopes.}.  A simple test to cross match
galaxies from SDSS with the RC3 catalogue shows that bright, extended
galaxies at low redshift can have large systematic errors in their SDSS
photometry.  This is likely due to the way that the SDSS algorithm treats
overlapping fields.  When we use SDSS to obtain photometry for extended
galaxies such as low-redshift BCGs, the effect of these photometry errors is
to underestimate the BCG's contribution to the overall stellar mass of the
cluster.  This problem is only significant for the low redshift X-ray groups
in the X-ray calibration sample, not for the clusters in the X-ray
calibration sample (which were chosen to sample the same redshift range as
our optically-selected LRG group/cluster sample) or our optically-selected
group and cluster sample.

To ensure more accurate stellar mass estimates for the X-ray groups and clusters, we first identify the BCG spectroscopically in SDSS, by ensuring that $|z_{\mathrm{sdss}}-z_{\mathrm{xray}}| \leq 0.001$, and then calculate the total K-band luminosity using the 2MASS survey\footnote{We take into account the systematic underestimation of 2MASS photometry by 20 per cent \citep[as discussed in][]{2003ApJ...591..749L}.}. We then use the Bell et al.\ prescription to calculate the BCG's stellar mass, and add it to the stellar masses of the satellites calculated using SDSS. We correct individual X-ray points for incompleteness, deprojection and apply the Gallazzi correction (Equation \ref{eq:bellcorr}). The X-ray points are plotted alongside our overall cluster sample result in Fig. \ref{cap:xraycomp}, and we see good consistency across the range of total masses with perhaps a slight inconsistency at the low mass end.  The remaining residual difference at low halo masses is plausibly due to the fact that we have corrected the luminosity of only the BCG (as opposed to all large, extended group galaxies) of the low-mass groups using 2MASS data .


\begin{figure*}
(a)
\includegraphics[width=0.4\textwidth]{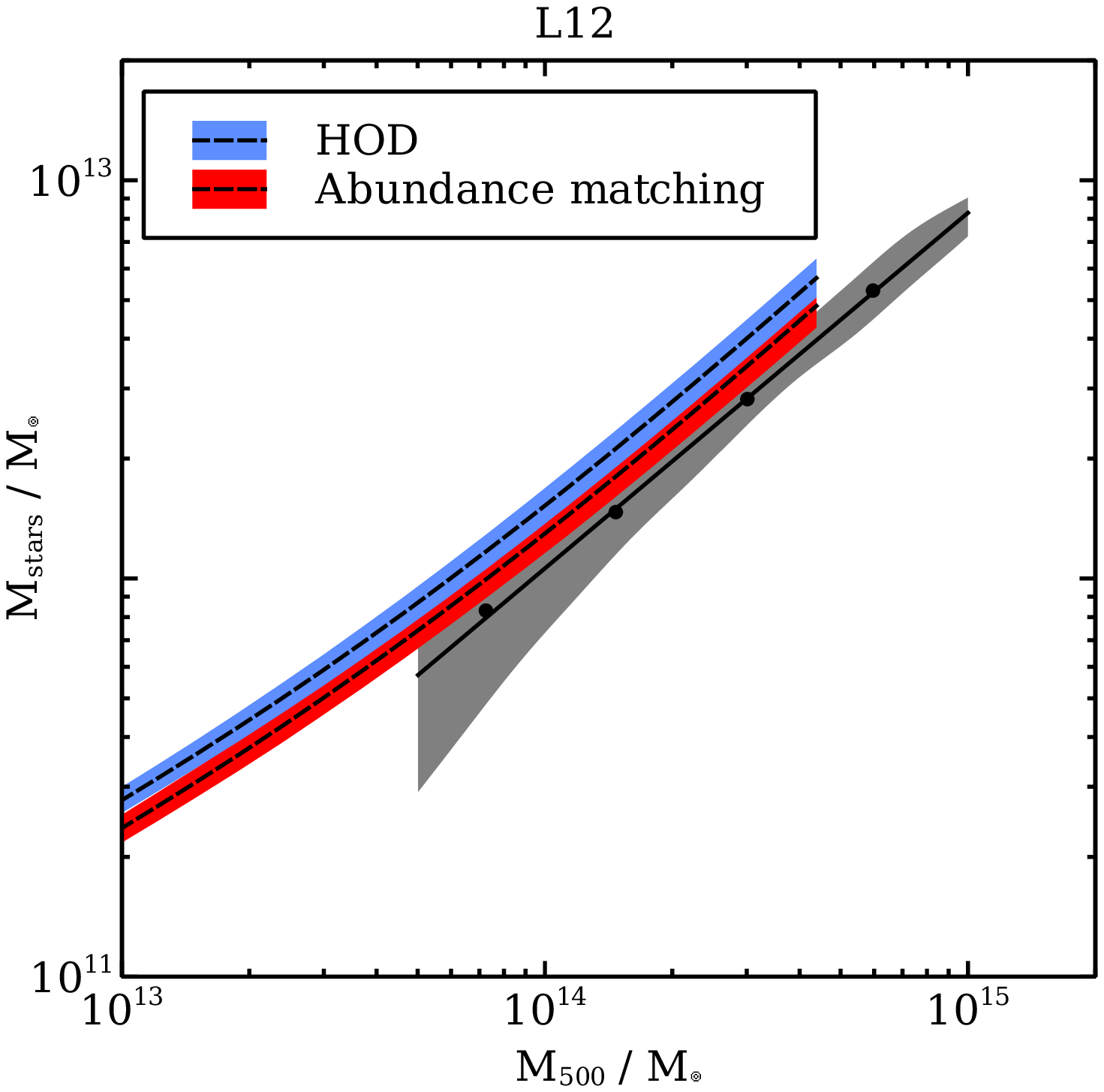}
\hspace{5mm}
(b)
\includegraphics[width=0.4\textwidth]{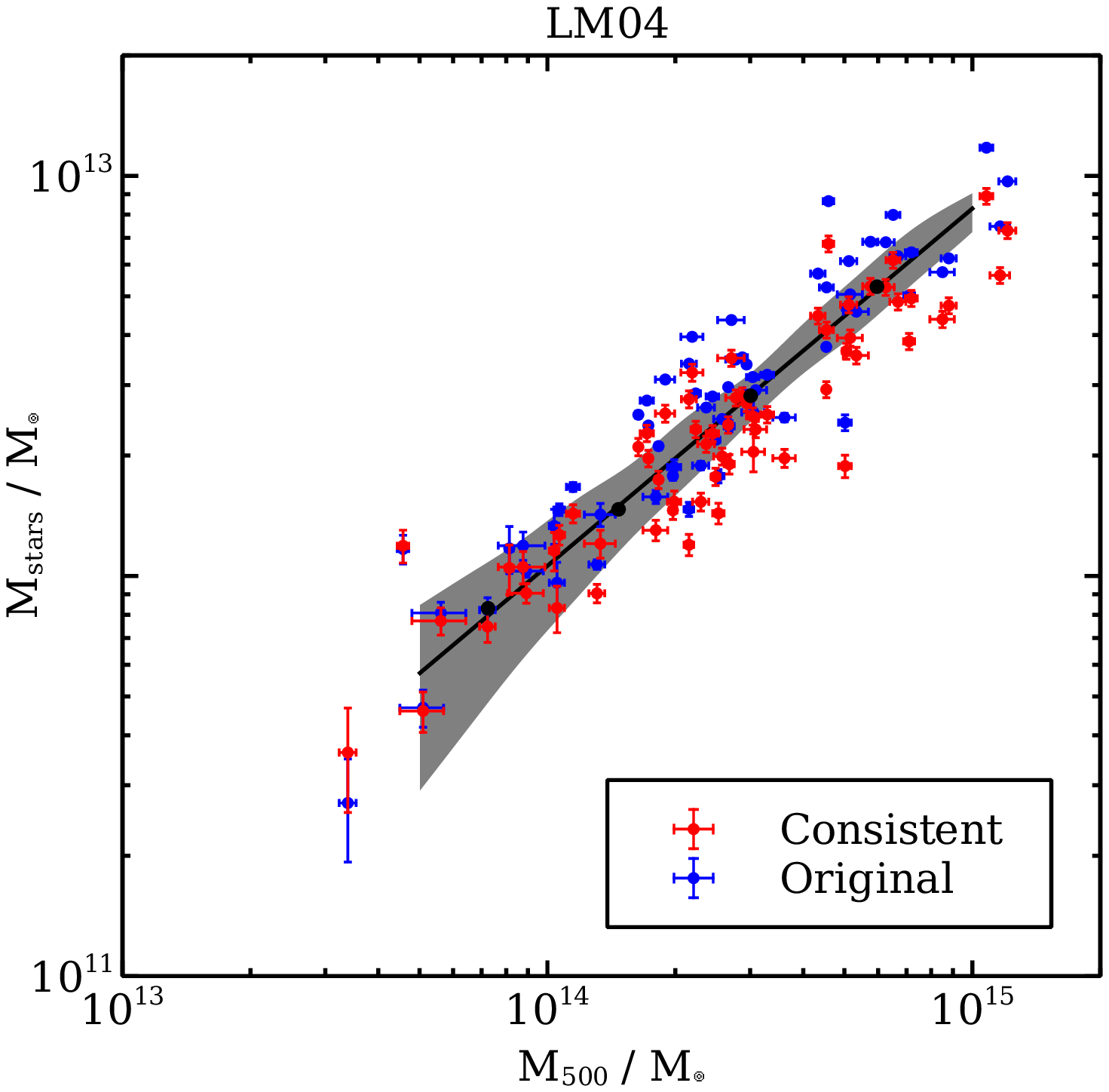}
\vspace{5mm}

(c)
\includegraphics[width=0.4\textwidth]{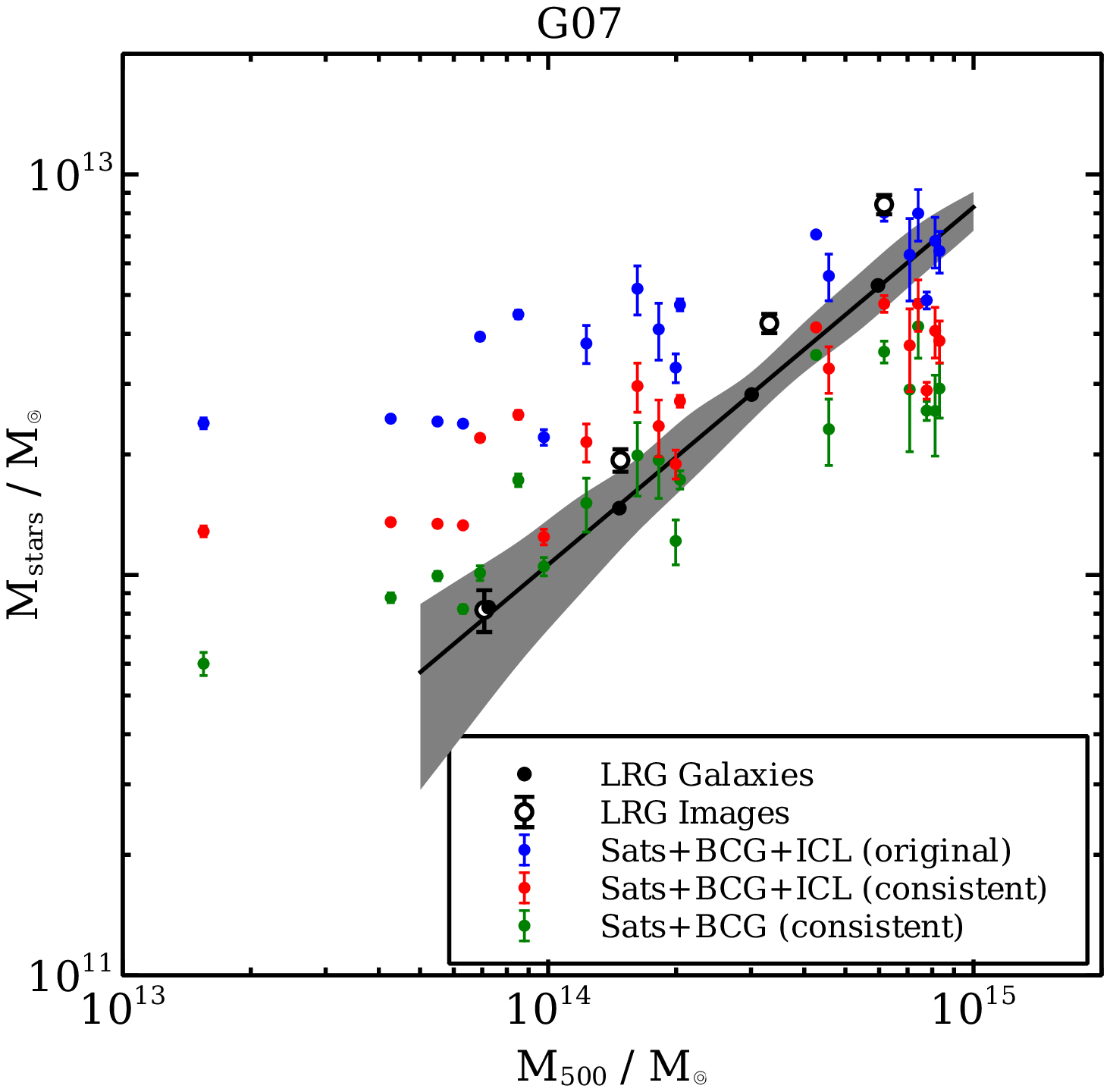}
\hspace{5mm}
(d)
\includegraphics[width=0.4\textwidth]{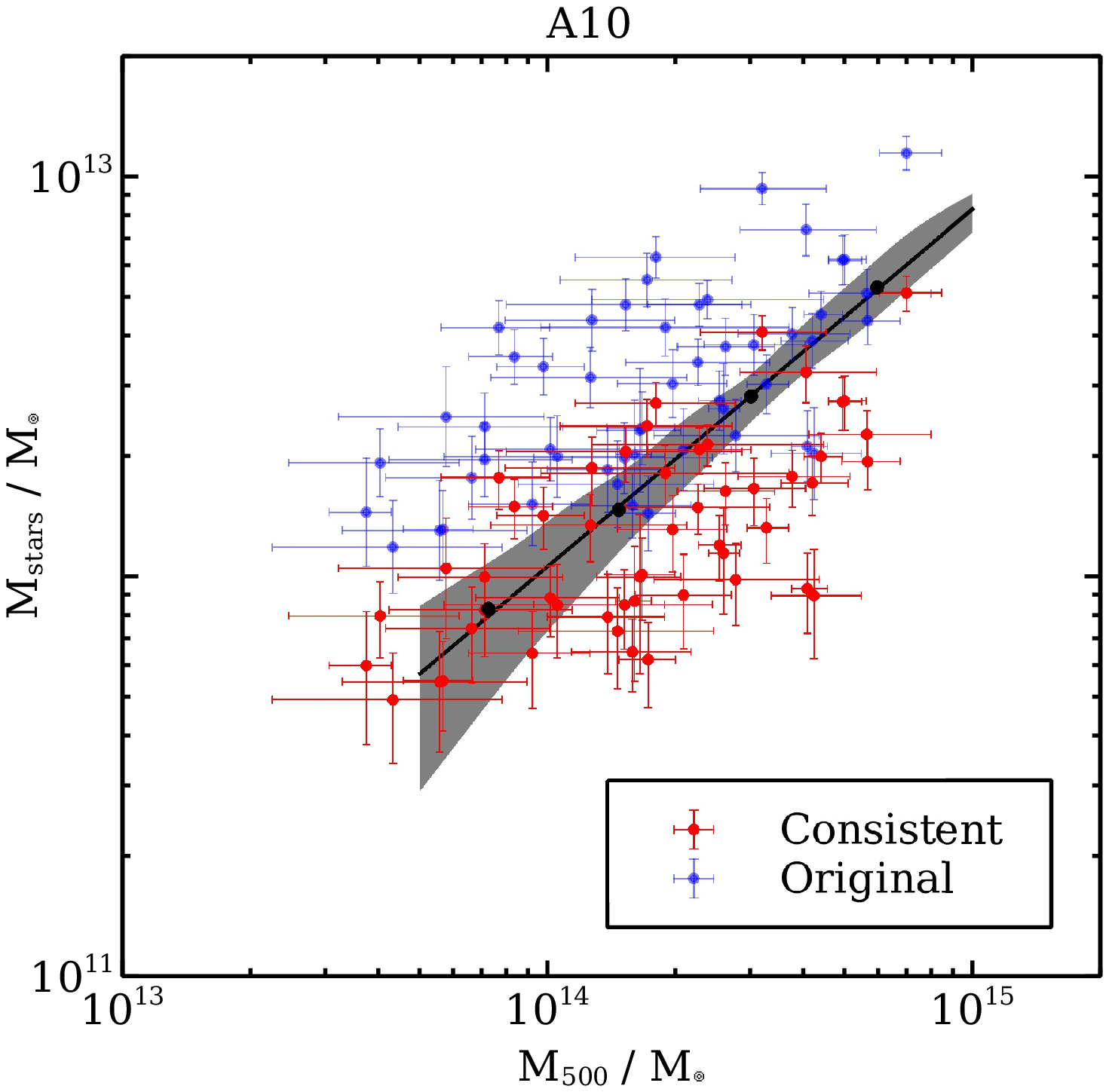}

\vspace{5mm}

\caption{A comparison of our stellar/total mass measurements with results
from other published work.  In each panel we compare our stacked result and
best-fit relation (black points and line) with a specific study from the
literature, and the grey region represents the $1\sigma$ error bar.  Panel
(a): Comparison with the abundance matching result of L12 (dashed line) and
corresponding error estimate (red region).  Panel (b): A comparison with the
X-ray cluster work of LM04, where the blue point show the original stellar
mass estimates, and the red points show the stellar mass estimates
calculated in a consistent way to this work.  Panel (c): A comparison with
the work of G07.  We split the sample by stellar mass into Satellites + BCG
(green points), and total stellar mass i.e.  Satellites + BCG + ICL (red
points).  We also include the original quoted stellar mass estimates in
blue, and include the image stacking points as empty circles.  Panel (d): A
comparison with A10.  Again the original stellar masses are in blue, and our
consistent comparison is in red.} \label{cap:mstarmhalo} \end{figure*}

\subsection{Comparison with previous work}\label{sec:comparison}

As discussed in Section 1, there are some significant differences in the
findings of previous studies on the stellar mass fractions of groups and
clusters and that this may be attributable, at least in part, to differences
in the adopted $M/L$ conversion (noting that assumption of a fixed $M/L$ is
better in the near-infrared than in optical bands).  Other potentially
relevant differences include whether the results were deprojected (i.e.,
presented within a sphere or a cylinder), if a correction was applied for
galaxies that were too faint to detect, and whether or not the ICL was
included, and how $M_{500}$ was estimated.  Below, we attempt to compare our
work with previous studies in a fair manner by adopting a consistent set of
$M/L$ values and consistently applying (or removing) corrections for
deprojection, the faint end of the luminosity function, and the ICL.  We
discuss whether differences in estimates of $M_{500}$ are relevant.  Fig. 
\ref{cap:mstarmhalo} compares the $M_{\rm star}-M_{500}$ relations from
previous studies, which we discuss in detail below.

\subsubsection{Leauthaud et al.\ (2012) - L12}

This study provides a prediction for $f_{\rm star}$ based on a statistical Halo Occupation Distribution (HOD) model which has been calibrated using joint constraints from lensing, clustering and the galaxy stellar mass function \citep{2011ApJ...738...45L}.  L12 also provide estimates of $M_{500}$ (and thus $f_{\rm star}$) based on the abundance matching (AM) technique \citep[details described in][]{2010ApJ...717..379B}, which links the stellar mass function from COSMOS to the halo mass function from cosmological simulations by assuming a monotonic relationship between stellar mass and halo mass and simply matching abundances.

To derive stellar masses, L12 fit 8 band photometry from COSMOS
\citep{2007ApJS..172....1S} using \citet{2003MNRAS.344.1000B} models
assuming a Chabrier IMF.  Their low-$z$ sample has a redshift range of $0.22
\leq z \leq 0.5$, which overlaps well with our sample.  As the HOD and AM
results are intrinsically 3-dimensional, no deprojection correction is
necessary.  Also, the depth of the COSMOS data allow L12 to measure galaxy
masses sufficiently far down the galaxy stellar mass function that the
correction for incompleteness (i.e., for faint galaxies they do not detect)
is expected to be negligible.  They do not, however, measure the
contribution from ICL.  We therefore directly compare the results of our
stacked galaxy catalog method (Section 3.1) to those of L12 results in Fig. 
\ref{cap:mstarmhalo} (a).

After applying the various mass, deprojection and missing flux corrections to our results, we find good agreement with the results of both methods employed by L12 in terms of the shape of the relation.  There is a 30 per cent offset in terms of normalisation with the HOD model predictions but, interestingly, only a $\sim$ 10 per cent offset from their AM results. There are a few possible explanations for the discrepancy between the HOD and AM models, as discussed in L12.  One possibility is that the additional constraints from galaxy-galaxy lensing and clustering help to better constrain total masses (the AM technique relies solely on the galaxy stellar mass function).  On the other hand, the HOD model of L12, at least in its current form, assumes that satellite galaxies trace the underlying dark matter, which we have shown in B12 to be a fairly poor approximation to the observed radial distribution of satellites in groups and clusters [i.e., observed satellite galaxies have a more extended (lower concentration) radial distribution than the dark matter].  This will bias the HOD total mass estimates, and therefore the inferred stellar mass fractions, at some level, but it is unclear if this bias alone can account for the difference in the HOD and AM results.  

\begin{figure} \includegraphics[width=\columnwidth]{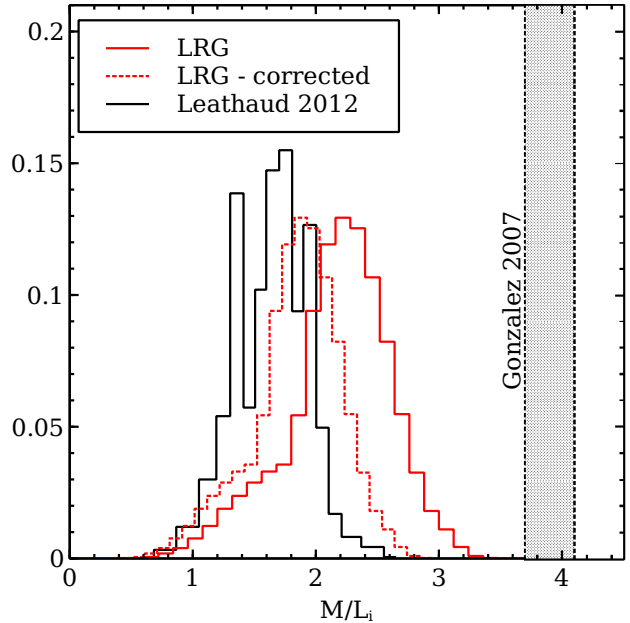}
\caption{A comparison of the SDSS $i$-band mass-to-light ratios in  
for the cluster sample (red), with those adopted by L12 and G07.  
The group/cluster sample mass-to-light ratios are obtained using the 
prescription of \citet{2003ApJS..149..289B}.  The dotted red line
shows the mass-to-light ratios after applying corrections described in Section
3.1. They seem to significantly improve the agreement with the findings of L12.} \label{cap:m2l_iband} \end{figure}

L12 have shown that differences in assumptions about the star formation
history and metallicity of the galaxies (which are input parameters into the
\citealt{2003MNRAS.344.1000B} models), as well as different assumptions
about the absorbing dust column, can lead to shifts in the stellar mass
estimates by up to $\sim 45$ per cent.  We have therefore performed an
additional test to see whether our \citet{2003ApJS..149..289B} stellar $M/L$
ratios are consistent with those adopted by L12.  We find quite good
agreement in the distribution of $i$-band mass-to-light ratios in passive
galaxies (once the mass correction in Equation \ref{eq:bellcorr} has been
applied - see Fig \ref{cap:m2l_iband}).  Therefore, differences in the
conversion from light to stellar mass do not appear to be the source of the
small offset with our results.  The offset is sufficiently small though (at
least with regards to the AM results), that it is conceivably due to
differences in the measured total light of galaxies (e.g., differences in
how the surface brightness profiles are extrapolated to large radii to
estimate the total light).

\subsubsection{Lin \& Mohr (2004) - LM04}

These authors present a series of papers
\citep{2003ApJ...591..749L,2004ApJ...610..745L,2004ApJ...617..879L} which
aim to investigate stellar mass fractions, K-band luminosity functions and
ICL in a sample of X-ray clusters with $0.02 \leq z \leq 0.09$, selected
from the X-ray flux-limited sample of \citet{2002ApJ...567..716R}.  Total
mass estimates are derived from using the X-ray mass - temperature relation
of \citet{2001A&A...368..749F}.  Although similar to the relation used to
calibrate our total masses \citep{2006ApJ...640..691V}, we recalculate the
total masses of LM04 using this more up-to-date relation for consistency
with our method.

LM04 use galaxy photometry from the 2MASS survey \citep{2006AJ....131.1163S}, and they calculate total K-band group and cluster luminosities within $r_{500}$ using a similar statistical background subtraction method to that described in Section \ref{sub:photostack}. A comparison of the LM04 stellar mass measurements with our result is shown in blue in Fig. \ref{cap:mstarmhalo} (b). There are however a number of subtle differences which need to be taken into account in order to ensure a fair comparison with our work. To reduce any uncertainty associated with different mass-to-light ratios and IMFs, we recalculate the LM04 stellar masses using our \citet{2003ApJS..149..289B} prescription (in the K-band). We use the fixed $M/L_{K}=0.7$ which does not change with total mass (see Table \ref{tab:brightscols}), and apply the \citeauthor{2003ApJS..149..289B} correction (Equation \ref{eq:bellcorr}), and the consistent comparison is shown in red.

LM04 also include a deprojection correction (similar to that described in Section \ref{sub:deproj}), which they state is of order 20-30 per cent. Our stellar masses are corrected to include \emph{all} missing flux (Section \ref{sub:missflux}), but LM04 correct for missing flux down to $K_{s}=-20$. In practice this makes very little difference, but we include an additional correction factor of 4 per cent (i.e. multiply all LM04 stellar masses by 1.04).

In general, we see very good agreement with this study across the group and cluster mass scales (Fig. \ref{cap:mstarmhalo} (b)). 

\subsubsection{Gonzalez et al.\ (2007) - G07}\label{sub:gonzcomp}

G07 focus on a sample of 23 nearby groups and clusters ($0.04 \leq z \leq 0.13$) observed with the Great Circle Camera \citep{1996PASP..108..104Z}.  The groups and clusters were selected to have a single dominant BCG with no signs of an ongoing merger \citep{2005ApJ...618..195G}.  Total masses are obtained using a X-ray hydrostatic mass - velocity dispersion relation, calibrated using a sample of 7 systems from \citet{2006ApJ...640..691V} with velocity dispersion estimates from the literature.

Completeness-corrected luminosities are obtained in the Johnson $I$-band by
using a statistical background subtraction method, which are then converted
to stellar masses using the mean mass-to-light ratio of
\citet{2006MNRAS.366.1126C}.  Interestingly, G07 find evidence for a
significant fraction of the total stellar mass residing in a diffuse ICL
component.  We show a comparison between the G07 stellar masses (blue
points) and our result in Fig.  \ref{cap:mstarmhalo} (c).  To perform a
consistent comparison, we use the \citet{2003ApJS..149..289B} prescription
to obtain $M/L_{I}$, and apply the mass correction in Equation
\ref{eq:bellcorr} to obtain stellar mass estimates (Fig. 
\ref{cap:mstarmhalo} (c)) We also include an deprojection correction of
\deprojection \ as this was not included in the G07 analysis.  We split the
G07 stellar masses to include estimates of stellar mass both with and
without the ICL component.

Focusing first on the comparison of the galaxies alone (i.e., BCG+satellites
but no ICL - compare green points with filled black circles), we find rough
agreement in terms of normalisation.  However, there is clear a difference
in the slope of the relations, such that G07 have significantly lower
stellar mass fractions at high total masses.  The reason(s) for this large
shape difference is unclear.  It is {\it not} due to differences in the
contribution of the ICL, as this component has been excluded in this
comparison.  It is also not likely due to a strong trend in the mean stellar
$M/L$ with total mass, $M_{500}$, as this varies by less than 10 per cent
over the mass range considered here (see Section 3.4).  Can differences in
total mass estimates account for the difference in the slopes of the
relations?  As noted above, G07 use a hydrostatic mass - velocity dispersion
relationship to estimate the total masses of their groups and clusters.  The
$\sigma$-$M_{500}$ relation adopted by G07 is very similar to that of
\citet{2008ApJ...672..122E}, and so by extension is consistent with our
observed relation (see Fig.  \ref{cap:mass_test}).  However, we have shown
there is a large amount of scatter in $\sigma$-$M_{500}$, which could potentially lead to large biases in the recovered $M_{\rm star}$-$M_{500}$ relation based on a small number of clusters (as in G07).  Alternatively, there may be a unknown systematic error in the derived luminosities of the most massive clusters, which biases them low.

Turning now to the contribution of the ICL to the total stellar mass
(compare red points with empty circles), in general agreement with G07 we
find that the contribution of the ICL (+ any mass associated with galaxies
that is not encapsulated by SDSS `model' magnutudes.) can be significant, up to
$\sim 40$ per cent.  
However, unlike G07 we do not find that the
contribution of the ICL decreases significantly with increasing total mass (in fact it is mildly rising with increasing total mass).
Again, it is unclear what the origin of the difference is.
 We note, however, that our results are in broad
agreement with the predictions of the cosmological simulations (e.g., \citealt{2004ApJ...607L..83M,2007ApJ...666...20P,2007MNRAS.377....2M,2010MNRAS.403..768H,2010MNRAS.406..936P}), which consistently find flat or mildly rising ICL fractions with increasing mass and a fractional contribution similar to what we measure.

\subsubsection{Andreon (2010) - A10}

The study of A10 makes use of a sample of 52 SDSS clusters with $0.03 \leq z
\leq 0.1$, whose masses are derived from a caustic method
\citep{1997ApJ...481..633D,2006AJ....132.1275R}.  They use a method similar
to our galaxy catalog method (i.e., neglect ICL) and, as in the case of G07,
A10 adopt a stellar mass-to-light ratio conversion from
\citet{2006MNRAS.366.1126C}.  However, unlike other authors, they provide a
Bayesian fit to the luminosity function of each cluster separately to
characterise the total luminosity.

We convert the $M_{200}$ estimates quoted by A10 into $M_{500}$ estimates by assuming the systems reside in NFW haloes with concentrations given by \citet{2008MNRAS.390L..64D}. We calculate estimates of $M_{500}^{\star}$, by propagating through the effects of a smaller aperture ($r_{500}$), and assuming that the satellites are distributed according to an NFW halo with $c_{\mathrm{sat}}=2.6$ (see B12). We present the `original' A10 $M_{\star}$,$M_{500}$ relation in Fig. \ref{cap:mstarmhalo} (d). However, in order to provide a consistent comparison with ours and other works we need to apply the Bell et al.\ prescription to calculate the stellar masses. We obtain the luminosities within $r_{500}$ for the A10 points by dividing the stellar masses by 3.8 as A10 have used a fixed mass-to-light ratio of 3.8 from \citet{2006MNRAS.366.1126C}. We then obtain the stellar masses using our method described in Section \ref{sub:photostack} and the comparison is included in Fig. \ref{cap:mstarmhalo} (d). A10 use only red galaxies in their analysis, and so we apply Equation \ref{eq:bellcorr} with $f_{\mathrm{red}}=1$. 

In an overall sense there is rough agreement (within the scatter) when a consistent comparison is made with A10.  A closer inspection reveals a mismatch in normalisation ($\sim$30) and that our relation is somewhat steeper than that of A10.  Interestingly, the scatter in the A10 data appears to be larger than in the studies we compared to above.

As the stellar mass measurements of A10 are derived from SDSS, we are able
to perform our own analysis of these same clusters.  Reassuringly, we find
that excellent agreement (with low scatter) is achieved at all masses with
the results of A10 if we adopt the {\it same aperture} as A10 (i.e., use
their $r_{200}$) and make a self-consistent comparison (i.e., adopting the
same stellar $M/L$, apply deprojection and completeness corrections).  This
agreement implies that the differences that are visible in the Fig. 
\ref{cap:mstarmhalo} (d) comparison are due to differences in the total mass
estimates, not in the stellar mass estimates (modulo the error introduced by
having a different aperture).  In particular, our total masses at the
intermediate- to high-mass end are slightly larger than what the caustic
measurements adopted by A10 imply.  The origin of this difference is not
clear, but we point out that the X-ray hydrostatic mass formalism we adopt
has been tested both against simulations and weak lensing observations with
the general consensus that the masses are accurate to 10-15 per cent within
$r_{500}$ (e.g.,
\citealt{2006MNRAS.369.2013R,2007ApJ...655...98N,2008MNRAS.384.1567M}). 
This is fully consistent with our findings in Section 2.1.

\section{Discussion and conclusions}

We have used the enormous power of the SDSS to perform a detailed statistical
analysis of the stellar mass fractions in groups and clusters in the mass range
$10^{13.7}-10^{15.0}$ $M_{\odot}$ at $0.15 < z < 0.4$, with a much larger sample
of systems than used previously. By stacking groups and clusters in bins of
total mass, we are able to obtain strong constraints on how the \emph{average}
stellar mass fraction changes with total mass. The total mass estimates
\citep[taken from an X-ray temperature derived mass and richness calibration
from][]{2012MNRAS.423..104B} are demonstrated to be accurate within $\approx$10
per cent on average using velocity dispersions and weak lensing. We provide two
estimates of the stacked stellar mass in each mass bin by making use of the
galaxy catalog data and images from SDSS-III, allowing us to quantify what is
the (maximum) fraction of the stellar mass contributed by any diffuse
intracluster light (ICL) component.

Our main results may be summarised as follows:
\begin{itemize}
\item{The stellar mass fraction (including ICL) within $r_{500}$ is typically
$0.013$ ($0.023$) assuming a Chabrier (Salpeter\footnote{As is standard, we
calculate stellar masses for a Salpeter IMF as a Chabrier IMF + 0.25 dex.}) IMF
and depends only very weakly on total mass (see the right panel of Fig.~\ref{cap:ourresult}).}
\item{The contribution of the ICL is significant, typically contributing
$20-40$ per cent of the total stellar mass.}
\item{Combining our results with gas mass fractions from the literature, we conclude that galaxy groups have baryon fractions that are well below the universal value.}
\end{itemize}

Many previous studies have attempted to measure the stellar mass fraction in
individual groups and clusters.  The comparison of our results with other
published works is an important part of this study and was a non-trivial
exercise, as different studies adopted different methods of calculating the
stellar mass from the luminosity, provide different corrections for missing
flux below the magnitude limit, and do not always provide a correction to
convert the projected stellar mass to a 3D stellar mass.  The ICL is also
treated differently in different works (it is generally ignored).  We have
compared with a range of published works and gone to great lengths to ensure
a fair comparison.  Most of the differences in the reported measurements can
be attributed to differences in the adopted stellar mass-to-light ratio and
differences in the halo mass estimates.  We are able to reconcile all of the
previous observational results we have compared to in this way.  The only
exception to this is in the comparison to \citet{2007ApJ...666..147G}, who
find significantly {\it lower} stellar mass fractions (with or without ICL)
in the most massive clusters compared to all other studies we have examined
when a consistent stellar mass-to-light ratio is adopted. The origin of this
difference remains unclear.

Our main result, that the stellar mass fraction depends very weakly on total mass, jibes well with the predictions of both semi-analytic models of
galaxy formation and full cosmological hydrodynamical simulations (e.g.,
\citealt{2005ApJ...625..588K,2006MNRAS.367.1641B,2006MNRAS.370..645B,
2010MNRAS.406..936P,2010MNRAS.406..822M}).  Note that while there are (sometimes
large) differences in the predicted normalisations of the $f_{\rm
star}$-$M_{500}$ relations from different models and simulations, to our
knowledge they all consistently predict similarly flat shapes.  This is not
surprising - as discussed in detail in \citet{2008MNRAS.385.1003B} it is
difficult to see physically how there could be a very steep relationship (as
claimed by both \citealt{2007ApJ...666..147G} and
\citealt{2010MNRAS.407..263A}), given that clusters are assembled (by mass)
primarily from the accretion of groups\footnote{It is perhaps even more
difficult to conceive of a physical scenario in which the {\it total baryon}
fractions of groups and clusters are very similar (i.e., do not depend on total
mass) yet are not equal to the universal mean ($\Omega_b/\Omega_m$), which is
what is implied at face value by the results of \citet{2007ApJ...666..147G} and
\citet{2010MNRAS.407..263A}.  In the limit where feedback is not energetic
enough to eject baryons from groups or clusters (and/or their high redshift
progenitors), the total baryon fraction ought to be close to the universal mean,
as simulations have consistently shown over the past two decades.  In the limit
where feedback {\it is} energetically sufficient to eject baryons, we expect the
ejection to be more efficient in low-mass systems due to their lower binding
energies, and thus the baryon fraction should increase with total mass.  Our
observations strongly suggest that the latter scenario is closer to the truth
(see the right panel of Fig.~\ref{cap:ourresult}).}.

With regards to the normalisation of the $f_{\rm star}$-$M_{500}$ relation,
there remain significant uncertainties due to our ignorance of the stellar
mass-to-light ratio.  In a given passband, the stellar mass-to-light ratio
depends on the stellar initial mass function (IMF), the age and metallicity
(distributions) of the stellar population, and the dust absorption column along
the line of sight. \citet{2012ApJ...746...95L} argue that non-IMF systematic
uncertainties can be as large as 45 per cent.  Adopting this estimate, then for
a Chabrier (Salpeter) IMF, we find $f_{\rm star} = 0.013^{+0.006}_{-0.004}$
($f_{\rm star} = 0.023^{+0.010}_{-0.007}$) for a typical mass of $M_{500} =
3\times10^{14}$ M$_\odot$.  For a universal baryon fraction $f_b \equiv
\Omega_b/\Omega_m = 0.165$ \citep{2013ApJS..208...19H}, this implies a (weakly
total mass-dependent) star formation efficiency of $f_{\rm star}/f_b =
8.3^{+3.7}_{-2.6}\%$ ($13.9^{+6.4}_{-4.3}\%$) in the case of a Chabrier (Salpeter) IMF.  
Thus, our results point to a relatively low star formation efficiency for groups
and clusters.  While our estimates are lower than some previous studies of
groups and clusters, they are reassuringly consistent with the {\it global} star
formation efficiency predicted by semi-analytic models that match the galaxy
luminosity function (e.g., \citealt{2006MNRAS.370..645B,2006MNRAS.365...11C}),
as well as the predictions of the abundance matching technique (e.g.,
\citealt{2012ApJ...746...95L}).

In the future we plan to study the individual contributions of the satellite galaxies, the BCG, and ICL to the total stellar mass.  The relative proportions of these components, as well as their spatial distributions, ought to constrain the efficiency of tidal stripping and dynamical friction of infalling galaxies and offer a means for improving the prescriptions of these processes in semi-analytic models.


\section*{Acknowledgements}

The authors thank the anonymous referee for a detailed and constructive report.
We would also like to thank Stefano Andreon, Matt Auger, Alexie Leauthaud, Sean McGee, Trevor Ponman, Eduardo Rozo, and Alastair Sanderson for valuable discussions. JMB acknowledges the award of a STFC research studentship. IGM acknowledges support from a STFC Advanced Fellowship.  This work has made extensive use of the \texttt{NumPy} and \texttt{SciPy} Numerical Python packages and the \texttt{Veusz} plotting package. Funding for the Sloan Digital Sky Survey (SDSS) has been provided by the Alfred P. Sloan Foundation, the Participating Institutions, the National Aeronautics and Space Administration, the National Science Foundation, the US Department of Energy, the Japanese Monbukagakusho, and the Max Planck Society. The SDSS website is http://www.sdss.org/. 
The SDSS is managed by the Astrophysical Research Consortium (ARC) for the Participating Institutions. The Participating Institutions are The University of Chicago, Fermilab, the Institute for Advanced Study, the Japan Participation Group, The Johns Hopkins University, Los Alamos National Laboratory, the Max Planck Institute for Astronomy (MPIA), the Max Planck Institute for Astrophysics (MPA), New Mexico State University, the University of Pittsburgh, Princeton University, the United States Naval Observatory, and the University of Washington.

\label{lastpage}


\appendix

\section{Quantifying covariance and scatter between mass bins}\label{sec:app}

The fact that the stellar mass estimates and the total mass calibration are both based on properties of the stellar population means there will be a degree of covariance between the two variables. It is important to model the impact of this covariance to ensure that our $M_{\star}$ vs $M_{500}$ plot is not entirely driven by correlated errors. 

We can model attempt to model this covariance by creating the characteristic error ellipses as a function of total mass. We use the following prescription:
\begin{enumerate}
\item Generate a cluster with $M_{500}$ equal to an input mass (different for each bin), and calculate the corresponding richness within 1 Mpc ($N_{1\mathrm{mpc}}$) by inverting the total mass - richness relation (Equation \ref{eq:masscalib}).
\item Using the projected NFW profile of \citet{1996A&A...313..697B}, calculate the corresponding richness within $r_{500}$ ($N_{r500}$) with a satellite concentration of $c_{\mathrm{sat}}=2.6$.
\item Then define a richness difference $\Delta n = N_{1\mathrm{mpc}} - N_{r500}$, which allows us to describe both the total mass and stellar mass according to the following model:

\begin{equation}
M_{500}\propto N_{1\mathrm{mpc}}^{1.4},\\
M_{\star}=\left( N_{1\mathrm{mpc}} - \Delta n  \right).\left< L_{r} \right> \left< M/L_{r} \right>
\label{eq:massscatter}
\end{equation}

\noindent where the mean luminosities and mass-to-light ratios are taken from Table \ref{tab:brightscols} for each bin.
\item Then draw an independent\footnote{Due to correlated errors between  between $M_{\star}$ and $M_{500}$ we need to apply scatter to both $N_{1\mathrm{mpc}}$ and $\Delta n$ independently.} Poisson sample from \emph{both} $N_{1\mathrm{mpc}}$ and $\Delta n$ (which also includes a background galaxy component), and recalculate the masses in Equation \ref{eq:massscatter}. The mean luminosity is drawn from a Gaussian distribution with mean of $3 \times 10^{10}$ $L_{\odot}$ (according to Table \ref{tab:brightscols}) and standard deviation\footnote{This is estimated by sampling from the luminosity function shown in Fig. \ref{cap:grouplfs}.} of $6 \times 10^{8}$ $L_{\odot}$. We now have a estimate of both $M_{\star}$ and $M_{500}$ with Poisson affects applied.
\item Repeat the above process 1000 times for each mass bin to generate the error ellipse distributions and fit with a 2D elliptical Gaussian function.
\end{enumerate}

\begin{figure}
\includegraphics[width=\columnwidth]{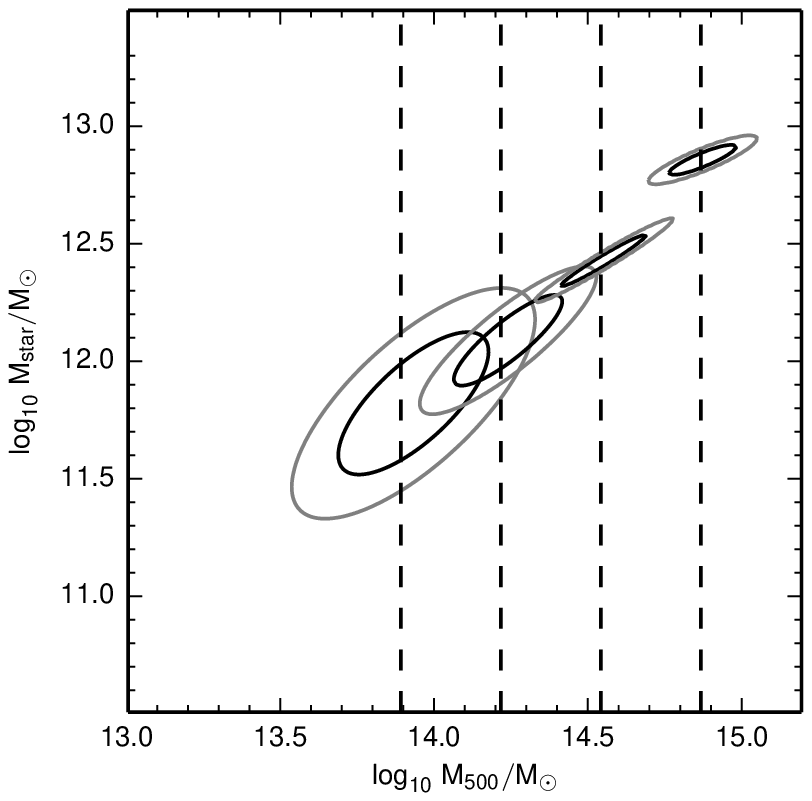}
\caption{A measure of the scatter between total mass bins due to the Poisson effects on the X-ray mass calibration. The vertical dashed lines show the input mass, and the black and grey ellipses show the 1 and 2 $\sigma$ contours once the Poisson effects have been applied. The tilt of the ellipses shows the significant covariance between the stellar mass and total mass which is most significant in the lowest mass bin. However, the lack of significant overlap between mass bins (especially at higher mass) shows that the observed stellar/total mass relationship is not caused by Poisson effects.}
\label{cap:scatter}
\end{figure}

The corresponding error ellipses (1 and 2 $\sigma$ contours) for each total mass bin are plotted in Fig. \ref{cap:scatter}. Despite a moderately large degree of covariance (tilt in the ellipse) in the lowest mass bin, this is reduced as the total mass increases.  Also, the degree of overlap between ellipses of clusters of different masses (particularly comparing the highest and lowest mass clusters) is also relatively modest, so we can be reasonably confident that any relation we deduce between $M_{\star}$ and $M_{500}$ is real.  We further note that this test was performed for the case of individual model clusters and so we can expect much smaller errors in the case of stacking large numbers of clusters of fixed richness.

Finally, that we see good correspondence between our stacked results and those deduced from our X-ray calibration sample (where we have an X-ray temperature estimate for each cluster and therefore an estimate of the halo mass which is independent of the stellar mass) confirms the robustness of the stacking results (see Fig. \ref{cap:xraycomp})

\end{document}